\documentclass[aps,prd,nofootinbib,showpacs,superscriptaddress,floatfix, preprint]{revtex4-1}
\pdfoutput=1
\usepackage[utf8]{inputenc}
\usepackage{amsmath,amssymb}
\usepackage{epsfig}
\usepackage{graphicx}
\usepackage[usenames,dvipsnames]{color}
\usepackage[colorlinks,citecolor=blue]{hyperref}
\usepackage{color}
\usepackage{subfigure}

\newcommand{\Tr}[1]{\text{Tr}\big[#1\big]}
\def\be{\begin{equation}}
\def\ee{\end{equation}}
\def\bsp#1\esp{\begin{split}#1\end{split}}

\def\bpm{\begin{pmatrix}}
	\def\epm{\end{pmatrix}}

\begin{document}
	\title{Phenomenological Study of Texture Zeros in Lepton Mass Matrices of Minimal Left-Right Symmetric Model}
	
	\author{Happy Borgohain}
	\email{happy@tezu.ernet.in}
	\affiliation{Department of Physics, Tezpur University, Napaam, Tezpur, Assam 784028, India}
	\author{Mrinal Kumar Das}
	\email{mkdas@tezu.ernet.in}
	\affiliation{Department of Physics, Tezpur University, Napaam, Tezpur, Assam 784028, India}
	\author{Debasish Borah}
	\email{dborah@iitg.ac.in}
	\affiliation{Department of Physics, Indian Institute of Technology Guwahati, Assam 781039, India}

	\begin{abstract}
		We consider the possibility of texture zeros in lepton mass matrices of the minimal left-right symmetric model (LRSM) where light neutrino mass arises from a combination of type I and type II seesaw mechanisms. Based on the allowed texture zeros in light neutrino mass matrix from neutrino and cosmology data, we make a list of all possible allowed and disallowed texture zeros in Dirac and heavy neutrino mass matrices which appear in type I and type II seesaw terms of LRSM. For the numerical analysis we consider those cases with maximum possible texture zeros in light neutrino mass matrix $M_{\nu}$, Dirac neutrino mass matrix $M_D$, heavy neutrino mass matrix $M_{RR}$ while keeping the determinant of $M_{RR}$ non-vanishing, in order to use the standard type I seesaw formula. The possibility of maximum zeros reduces the free parameters of the model making it more predictive. We then compute the new physics contributions to rare decay processes like neutrinoless double beta decay, charged lepton flavour violation. We find that even for a conservative lower limit on left-right symmetry scale corresponding to heavy charged gauge boson mass 4.5 TeV, in agreement with collider bounds, for right-handed neutrino masses above 1 GeV, the new physics contributions to these rare decay processes can saturate the corresponding experimental bound. 
	\end{abstract}
	\maketitle

	\section{Introduction}
	The fact that neutrinos have non-zero but tiny masses and large mixing has been well established by several neutrino experiments \cite{Fukuda:2001nk, Ahmad:2002jz, Ahmad:2002ka, Abe:2008aa, Abe:2011sj, Abe:2011fz, An:2012eh, Ahn:2012nd, Adamson:2013ue} during the last two decades. For a review of neutrino mass and mixing, please see \cite{Mohapatra:2005wg, Tanabashi:2018oca}. Among the above-mentioned experiments, the relatively recent ones like T2K \cite{Abe:2011sj}, 
	Double Chooz \cite{Abe:2011fz}, Daya Bay \cite{An:2012eh}, RENO \cite{Ahn:2012nd} and MINOS \cite{Adamson:2013ue} experiments have not only confirmed the results from earlier experiments but also discovered the non-zero reactor mixing angle $\theta_{13}$. For a recent global fit of neutrino oscillation data, we refer to \cite{deSalas:2017kay, Esteban:2018azc}. The latest global fit shows that a few details of the light neutrinos are yet to be determined experimentally. They are namely, the Dirac CP phase, octant of atmospheric mixing angle and the ordering of light neutrinos: normal ordering (NO) or inverted ordering (IO). Also, the nature of neutrinos (Dirac or Majorana) remains unknown at oscillation experiments. If neutrinos are Majorana fermions, there arise two more CP phases known as Majorana CP phases, which can not be determined by oscillation experiments and have to be probed at alternative experiments. Apart from neutrino oscillation experiments, the neutrino sector is constrained by the data from cosmology as well. For example, the latest data from the Planck mission constrain the sum of absolute neutrino masses $\sum_i \lvert m_i \rvert < 0.12$ eV \cite{Aghanim:2018eyx}.

	Although we have significant experimental observations related to the neutrino sector except for the above-mentioned unknowns, the dynamical origin of light neutrino masses and their mixing is still a mystery. The standard model (SM) of particle physics, which gives a successful description of all fundamental particles and their interactions (except gravity) can not explain the lightness of neutrinos. The Higgs field in the SM which is responsible for generating masses to all known particles do not have coupling to neutrinos as the right-handed (RH) neutrinos are absent. One can generate a light Majorana mass term for light neutrinos in the SM through the dimension five Weinberg operator \cite{Weinberg:1979sa} of type $(L L H H)/\Lambda$ with the introduction of an unknown cutoff scale $\Lambda$. Several beyond standard model (BSM) proposals have been put forward which can provide a dynamical origin of such operators in a renormalizable theory. This is typically achieved in the context of seesaw models where a seesaw between the electroweak scale and the scale of newly introduced fields decide the smallness of neutrino masses. Popular seesaw models can be categorized as type I seesaw \cite{Minkowski:1977sc, GellMann:1980vs, Mohapatra:1979ia, Schechter:1980gr}, type II seesaw \cite{Mohapatra:1980yp, Lazarides:1980nt, Wetterich:1981bx, Schechter:1981cv, Brahmachari:1997cq}, type III seesaw \cite{Foot:1988aq} among others like \cite{Ma:1998dn,Mohapatra:1986bd}.

	One very popular BSM scenario is the framework of the left-right symmetric model (LRSM) \cite{Pati:1974yy, Mohapatra:1974hk,%
		Mohapatra:1974gc, Senjanovic:1975rk, Mohapatra:1977mj, Senjanovic:1978ev,%
		Mohapatra:1980qe, Lim:1981kv, Gunion:1989in,Deshpande:1990ip, FileviezPerez:2008sr} where the gauge symmetry of the SM is extended to $ \rm SU(3)_c\times SU(2)_L\times SU(2)_R\times U(1)_{B-L}$ so that the right-handed fermions (which are singlet in SM) can form doublets under the new $SU(2)_R$. This not only makes the inclusion of right-handed neutrino automatic, but also puts the left and right-handed fermions on equal footing. If we also incorporate an additional discrete left-right symmetry to ensure that the theory is invariant under $SU(2)_L \leftrightarrow SU(2)_R$. So the model can explain the origin of parity violation in weak interaction by considering a parity symmetric theory at high energy scale where the corresponding gauge symmetry breaks spontaneously leading to the parity-violating SM at low energy. In the minimal LRSM, the light neutrino masses arise naturally from a combination of type I and type II seesaw. It should be noted that the idea of combining type I and type II seesaw mechanisms for light neutrino masses was pursued in several earlier works too, for example, \cite{Ioannisian:1994nx, Bamert:1994vc, Antusch:2004xd}. The gauge symmetry, as well as the particle content of minimal LRSM, can also be accommodated within popular grand unified theory (GUT) models like $SO(10)$. Apart from this, another interesting motivation for this model is its verifiability. A TeV scale LRSM can have very interesting signatures which are being looked at colliders \cite{Aaboud:2017efa,Aaboud:2017yvp,Sirunyan:2016iap,Khachatryan:2016jww,Sirunyan:2018mpc}. There also exist different other phenomenological consequences which can be probed at experiments in both energy as well as intensity frontiers.

	Typical seesaw models in the absence of specific flavour symmetries usually predict a very general structure of light neutrino mass matrix which can always be fitted to the observed data due to the presence of many free parameters. The same is true in LRSM as well. However, if the theory has a well-motivated underlying symmetry that gives rise 
	to a very specific structure of the neutrino mass matrix, the number of free parameters can be significantly reduced. In such a case, we can have very specific predictions for light neutrino parameters like CP phase, octant of atmospheric mixing angle, mass ordering which can be tested at ongoing experiments. Here we consider such a possibility where an underlying symmetry can restrict the mass matrix to have non-zero entries only at certain specific locations. 
	Such scenarios are more popularly known as zero texture models, a nice summary 
	of which within three neutrino framework can be found in the review article~\cite{Ludl:2014axa} \footnote{Also see \cite{Xing:2002ta, Singh:2016qcf, Ahuja:2017nrf, Borah:2015vra, Kalita:2015tda} for texture related works in different contexts.}. In the diagonal charged lepton basis, if the light neutrino mass matrix has some textures, the corresponding constraints can be solved to find the light neutrino parameter space that satisfies them. Depending on the viability of this parameter space in view of the latest neutrino oscillation data, one can discriminate between different textures. Also, the allowed textures often predict non-trivial values for unknown parameters that can be tested at different experiments. It has already been shown in earlier works that in the diagonal charged lepton basis, not more than two zeros are allowed in the light neutrino mass matrix. While all six possible one zero texture $(^6C_n, n=1)$ are allowed, among the fifteen possible two zero textures, only six were found to be allowed after incorporating both neutrinos as well as cosmology data \cite{Meloni:2014yea,Fritzsch:2011qv,Alcaide:2018vni,Zhou:2015qua,Bora:2016ygl,Borgohain:2018lro}. Since in LRSM, several mass matrices play a role in generating light neutrino mass matrix due to the combination of type I and type II seesaw, the requirement of getting the allowed texture zeros in light neutrino mass matrix can constrain the texture zeros of all other mass matrices in the lepton sector namely, the Dirac neutrino mass $M_D$ and heavy neutrino mass $M_{RR}$. Making a list of all these possibilities while classifying the allowed and disallowed ones is the primary goal of this work\footnote{Please see \cite{Borah:2016xkc, Borah:2017azf, Sarma:2018bgf} and references therein for texture zero works in $3+1$ neutrino scenarios and \cite{Nath:2016mts} for related phenomenological study of texture zeros in all relevant lepton mass matrices of a particular seesaw model.}. We not only make such a list considering all possibilities of texture zeros but also perform a numerical analysis for one zero and two zero light neutrino textures as well as a scenario where other mass matrices involved in the seesaw can have a maximum number of zeros. To be more specific, for our numerical analysis, we considered five zero textures in $M_D$ and four zero textures in $M_{RR}$, keeping the rank of the latter three. Out of 378 total possibilities belonging to this list, we find that 189 are allowed from light neutrino data, out of which 109 give rise to two zero textures in light neutrino mass matrix. The case for a maximum number of zeros is particularly chosen due to their more predictive nature. We not only find the correlations among light neutrino parameters, but also find the new physics contribution to other interesting processes like neutrinoless double beta decay (NDBD) and charged lepton flavour violation (CLFV). As these processes are being probed at several experiments, this study points out the possibility of probing such scenarios at those experiments. Such aspects of probing LRSM can be complementary to the ongoing collider searches mentioned earlier.

	This paper is organized as follows. In section \ref{sec2}, we review the LRSM with its particle content and mass spectrum followed by the details of the texture structures of the Dirac and Majorana mass matrices in section \ref{sec2a}. We then summarize the contributions to NDBD and CLFV in LRSM in section \ref{sec3}, \ref{sec3a} respectively. We discuss our numerical analysis and results in section \ref{sec4} and then finally conclude in section \ref{sec5}.

	\section{Minimal Left-Right Symmetric Model}
	{\label{sec2}}
	As mentioned before, the left-right symmetric model is a very well motivated and widely studied extension of the SM with an enlarged gauge symmetry based on $ \rm SU(3)_c\times SU(2)_L\times SU(2)_R\times U(1)_{B-L}$ \cite{Pati:1974yy, Mohapatra:1974hk,%
		Mohapatra:1974gc, Senjanovic:1975rk, Mohapatra:1977mj, Senjanovic:1978ev,%
		Mohapatra:1980qe, Lim:1981kv,Gunion:1989in, Deshpande:1990ip, FileviezPerez:2008sr}. The theory removes the disparity between the left and right-handed fields by considering the right-handed fields to be doublet under the additional $SU(2)_R$ keeping the right sector couplings same as the left one by left-right symmetry. Therefore, the fermion field content of the minimal LRSM can be written as
	\begin{equation}\label{eq1}
	\rm Q_L=\left[\begin{array}{c}
	u_L\\
	d_L
	\end{array}\right]\equiv \left(3,2,1,\frac{1}{3}\right),
	Q_R=\left[\begin{array}{c}
	u_R\\
	d_R
	\end{array}\right]\equiv \left(3,1,2,\frac{1}{3}\right) 
	\end{equation}
	\begin{equation} \label{eq2}         
	l_L=\left[\begin{array}{c}
	\nu_L\\
	e_L
	\end{array}\right]\equiv \left(1,2,1,-1\right),
	l_R=\left[\begin{array}{c}
	\nu_R\\
	e_R
	\end{array}\right]\equiv \left(1,1,2,-1\right)                   
	\end{equation}
	where the numbers in brackets represent the quantum numbers under the the gauge group $ \rm SU(3)_c\times SU(2)_L\times SU(2)_R\times U(1)_{B-L}$. The Higgs sector of the minimal LRSM consists of two $SU(2)_L$ triplets $\Delta_{L,R}$ and a bi-doublet $\phi$ given by
	\begin{equation}\label{eq3}
	\Phi=\left[\begin{array}{cc}
	\phi_1^0 & \phi_1^+\\
	\phi_2^- & \phi_2^0
	\end{array}\right]\equiv \left( \phi_1,\widetilde{\phi_2}\right), \Delta_{L,R}=\left[\begin{array}{cc}
	{\delta_\frac{L,R}{\sqrt{2}}}^+ & \delta_{L,R}^{++}\\
	\delta_{L,R}^0 & -{\delta_\frac{L,R}{\sqrt{2}}}^+ .
	\end{array}\right],
	\end{equation}
	with the quantum numbers $ \rm \Phi(1,2,2,0)$ and $\Delta_L(1,3,1,2)$, $\Delta_R(1,1,3,2)$ respectively.

	The relevant Yukawa Lagrangian giving masses to the three generations of leptons  is given by,
	\begin{equation}\label{eq4}
	\mathcal{L}=h_{ij}\overline{l}_{L,i}\Phi l_{R,j}+\widetilde{h_{ij}}\overline{l}_{L,i}\widetilde{\Phi}l_{R,j}+f_{L,ij}{l_{L,i}}^TCi\sigma_2\Delta_L l_{L,j}+f_{R,ij}{l_{R,i}}^TCi\sigma_2\Delta_R l_{R,j}+{\rm h.c},
	\end{equation}
	where the indices $i,j=1,2,3$ represent the family indices for the three generations of fermions. $C=i\gamma_2\gamma_0$ is the charge conjugation operator, $\widetilde{\Phi}=\tau_2\phi^*\tau_2$ and $\gamma_{\mu}, \tau_2$ are the Dirac and Pauli matrices respectively. Discrete left-right symmetry ensures the equality of Majorana Yukawa couplings $f_L = f_R$ apart from the equality of gauge couplings of $SU(2)_{L,R}$ sectors $g_L = g_R$. The scalar potential $V_{\text{scalar}}$ is
	given by
	{\small \be\bsp
		V_{\text{scalar}} & = -\mu_1^2 \Tr{\Phi^\dagger\Phi}
		- \mu_2^2\Tr{\Phi^\dagger\tilde{\Phi} + \tilde{\Phi}^\dagger\Phi}
		- \mu_3^2\Tr{\Delta_L^\dagger\Delta_L + \Delta_R^\dagger\Delta_R}
		+ \lambda_1\Big(\Tr{\Phi^\dagger\Phi}\Big)^2
		\\ &\
		+ \lambda_2\Big\{ \Big(\Tr{\Phi^\dagger\tilde{\Phi}}\Big)^2
		+ \Big(\Tr{\tilde{\Phi}^\dagger\Phi}\Big)^2\Big\}
		+ \lambda_3 \Tr{\Phi^\dagger\tilde{\Phi}}\Tr{\tilde{\Phi}^\dagger\Phi}
		+ \lambda_4\Tr{\Phi^\dagger\Phi}\Tr{\Phi^\dagger\tilde{\Phi}
			+ \tilde{\Phi}^\dagger\Phi} \\
		&\
		+ \rho_1\Big\{\Big(\Tr{\Delta_L^\dagger\Delta_L}\Big)^2
		+ \Big(\Tr{\Delta_R^\dagger\Delta_R}\Big)^2\Big\}
		+ \rho_2\Big\{\Tr{\Delta_L\Delta_L}\Tr{\Delta_L^\dagger\Delta_L^\dagger}
		+ \Tr{\Delta_R\Delta_R}\Tr{\Delta_R^\dagger\Delta_R^\dagger}\Big\}
		\\&\
		+ \rho_3\Tr{\Delta_L^\dagger\Delta_L}\Tr{\Delta_R^\dagger\Delta_R}
		+ \rho_4\Big\{\Tr{\Delta_L\Delta_L}\Tr{\Delta_R^\dagger\Delta_R^\dagger}
		+ \Tr{\Delta_L^\dagger\Delta_L^\dagger}\Tr{\Delta_R\Delta_R}
		\\&\
		+ \alpha_1\Tr{\Phi^\dagger\Phi}
		\Tr{\Delta_L^\dagger\Delta_L + \Delta_R^\dagger\Delta_R}
		+ \Big\{
		\alpha_2 \Big(\Tr{\Phi^\dagger\tilde{\Phi}}\Tr{\Delta_L^\dagger\Delta_L}
		+ \Tr{\tilde{\Phi}^\dagger\Phi}\Tr{\Delta_R^\dagger\Delta_R}\Big)
		+ \text{h.c.}\Big\}   \\&\
		+ \alpha_3 \Tr{\Phi\Phi^\dagger\Delta_L\Delta_L^\dagger
			+ \Phi^\dagger\Phi\Delta_R\Delta_R^\dagger}
		+ \beta_1 \Tr{\Phi^\dagger\Delta_L^\dagger\Phi\Delta_R
			+ \Delta_R^\dagger\Phi^\dagger\Delta_L\Phi}
		+ \beta_2\Tr{\Phi^\dagger\Delta_L^\dagger\tilde{\Phi}\Delta_R
			+ \Delta_R^\dagger\tilde{\Phi}^\dagger\Delta_L\Phi}
		\\ &\
		+ \beta_3\Tr{\tilde{\Phi}^\dagger\Delta_L^\dagger\Phi\Delta_R
			+ \Delta_R^\dagger\Phi^\dagger\Delta_L\tilde{\Phi}} \ ,
		\esp\label{appeneq2}\ee}
	where we have introduced scalar mass parameters $\mu_i$ and quartic scalar
	interaction strengths $\lambda_i$, $\rho_i$, $\alpha_i$ and $\beta_i$. In the symmetry breaking pattern, the neutral component of the Higgs triplet $\Delta_R$ acquires a vacuum expectation value (VEV) to break the gauge symmetry of the LRSM into that of the SM and then to the $U(1)$ of electromagnetism by the VEV of the neutral components of Higgs bidoublet $\Phi$:
	$$ SU(2)_L \times SU(2)_R \times U(1)_{B-L} \quad \underrightarrow{\langle
		\Delta_R \rangle} \quad SU(2)_L\times U(1)_Y \quad \underrightarrow{\langle \Phi \rangle} \quad U(1)_{\rm em}.$$
	The VEVs  of the neutral components of the Higgs fields can be denoted as
	\be
	\langle \phi^0_{1,2} \rangle = \frac{k_{1,2}}{\sqrt{2}}
	\qquad\text{and}\qquad
	\langle \Delta^0_{L, R} \rangle = \frac{v_{L,R}}{\sqrt{2}}\ ,
	\ee
	where the VEV's $k_1, k_2$ satisfy the VEV of the SM namely, $v_{\rm SM} =
	\sqrt{k_1^2+k_2^2} \approx 246$~GeV. The VEV $v_L$ which plays a significant role in neutrino mass mechanism is generated after the electroweak symmetry breaking due to the following induced VEV relation
	\begin{equation}\label{eq5}
	\langle \Delta_L \rangle=v_L=\frac{\gamma v^2_{\rm SM}}{v_R}.
	\end{equation}
	Here, $\gamma$ is a dimensionless parameter given by \cite{Deshpande:1990ip}
	\begin{equation}\label{eq10}
	\gamma=\frac{\beta_1k_1k_2+\beta_2{k_1}^2+\beta_3{k_2}^2}{(2\rho_1-\rho_3)k^2}.
	\end{equation}
	In order to satisfy the electroweak precision test constraints, $v_L$ should be
	smaller than 2~GeV~\cite{Agashe:2014kda}, and the above breaking pattern of
	gauge symmetry enforces $v_R$ to be much greater than $k_{1,2}$.

	The $6\times6$ neutrino
	mass matrix is then given, in the $(\nu_L, \nu_R)$ gauge eigenbasis, by
	\be
	M = \bpm \sqrt{2} fv_L & M_D \\ M^T_D & M_R \epm \ =\bpm M_{LL} & M_D \\ M^T_D & M_{RR} \epm \ 
	\ee
	Assuming $M_{LL}\ll M_D\ll M_R$, the light neutrino mass after symmetry breaking is generated within a type I+II seesaw as,
	\begin{equation}\label{eq9}
	\rm M_\nu= {M_\nu}^{I}+{M_\nu}^{II}
	\end{equation}
	\begin{equation}\label{eq6}
	M_\nu=M_{LL}-M_D{M_{RR}}^{-1}{M_D}^T
	=\sqrt{2}v_Lf_L-\frac{v^2_{\rm SM}}{\sqrt{2}v_R}h_D{f_R}^{-1}{h_D}^T,
	\end{equation}
	\begin{equation}\label{eq7}
	M_D=\frac{1}{\sqrt{2}}(k_1h+k_2\widetilde{h}), M_{LL}=\sqrt{2}v_Lf_L, M_{RR}=\sqrt{2}v_Rf_R,
	\end{equation}
	\begin{equation}\label{eq8}
	\rm h_D=\frac{(k_1h+k_2\widetilde{h})}{\sqrt{2} v_{\rm SM}}.
	\end{equation}
	$M_D$, $M_{LL}$ and $M_{RR}$ being the Dirac neutrino mass matrix, left-handed and right-handed
	Majorana mass matrix respectively. The first and second terms in equation (\ref{eq7}) correspond to type II seesaw and type I seesaw contributions respectively. 
	
	The $6\times 6$ neutral lepton mass matrix can be diagonalized by a $6\times 6$ unitary matrix, as follows,
	\begin{equation}\label{eq14}
	\mathcal{V}^T M \mathcal{V}=\left[\begin{array}{cc}
	\widehat{M_\nu}&0\\
	0&\widehat{M}_{RR}
	\end{array}\right],
	\end{equation}
	where, $\mathcal{V}$ represents the diagonalizing matrix of the full neutrino mass matrix, $M$, $\widehat{M_\nu}= {\rm diag}(m_1,m_2,m_3)$, with $m_i$ being the light neutrino masses and 
	$\widehat{M}_{RR}= {\rm diag}(M_1,M_2,M_3)$, with $M_i$ being the heavy right-handed neutrino masses. $\mathcal{V}$ is thus represented as,
	\begin{equation}\label{eq15}
	\mathcal{V}=\left[\begin{array}{cc}
	U&S\\
	T&V
	\end{array}\right] \approx \left[\begin{array}{cc}
	1-\frac{1}{2}RR^{\dagger}&R\\
	-R^\dagger&1-\frac{1}{2}R^\dagger R
	\end{array}\right] \left[\begin{array}{cc}
	V_\nu&0\\
	0&V_R
	\end{array}\right],
	\end{equation}
	where, R describes the left-right mixing and given by,
	\begin{equation}\label{eq16}
	R=M_D M^{-1}_{RR}+\mathcal{O} (M^3_D{(M^{-1}_{RR})}^3).
	\end{equation}
	The matrices U, V, S and T are as follows,
	\begin{equation}\label{eq17}
	U=\left[1-\frac{1}{2}M_DM^{-1}_{RR}{(M_DM^{-1}_{RR})}^\dagger\right]V_\nu,V=\left[1-\frac{1}{2}{(M_DM^{-1}_{RR})}^\dagger M_DM^{-1}_{RR}\right]V_R
	\end{equation}
	\begin{equation}\label{eq18}
	S=M_DM^{-1}_{RR}V_R,T=-(M_DM^{-1}_{RR})^\dagger V_\nu.
	\end{equation}

	The gauge boson mass spectra can be found similarly. The left-right gauge boson mixing is given by
	\begin{equation}\label{eq19}
	\left[\begin{array}{cc}
	W^{\pm}_L\\
	W^{\pm}_R
	\end{array}\right]= \left[\begin{array}{cc}
	\cos \xi & \sin\xi e^{i\alpha}\\
	-\sin \xi e^{-i\alpha} & \cos\xi
	\end{array}\right]\left[\begin{array}{cc}
	W^{\pm}_1\\
	W^{\pm}_2
	\end{array}\right] ,
	\end{equation}
	with the mixing parameter $\xi$ represented by
	\begin{equation}\label{eq20}
	\tan 2\xi= -\frac{2 k_1 k_2}{v^2_R-v^2_L}.
	\end{equation}
	Without any loss of generality, we make
	use of rotation in the $SU(2)_L\times SU(2)_R$ space so that only one of the neutral components of the Higgs bidoublet acquires a large vacuum expectation value, $k_1\approx v_{\rm SM}$ and $k_2\approx0$. This corresponds to negligible mixing $\xi$.
	
	Under those assumptions, we
	neglect all contributions to the gauge boson masses that are proportional to
	$v_L$, so that these masses approximatively read
	\be\bsp
	M^2_{W_L}= \frac{g^2}{4} k^2_1\ ,
	\quad\quad
	M^2_{W_R} = \frac{g^2}{2}v^2_R\ ,
	\quad\quad
	M^2_Z = \frac{g^2 k^2_1}{4\cos^2{\theta_W}}
	\Big(1-\frac{\cos^2{2\theta_W}}{2\cos^4{\theta_W}}\frac{k^2_1}{v^2_R}\Big)\ ,
	\quad\quad
	M^2_{Z'} = \frac{g^2 v^2_R \cos^2{\theta_W}}{\cos{2\theta_W}} \ ,
	\esp\label{Mgauge}\ee
	with $\theta_W$ indicating the weak mixing angle.
	
	Under these assumptions, the Dirac neutrino mass matrix is 
	\begin{equation}
	M_D=\frac{1}{\sqrt{2}}(k_1h)
	\end{equation}
	while the charged lepton mass matrix is 
	\begin{equation}
	M_l=\frac{1}{\sqrt{2}}(k_1\widetilde{h})
	\end{equation}
	which points out the freedom in choosing $M_l$ and $M_D$ as we do in the subsequent sections.
	
	\section{Texture Zeros in Lepton Mass Matrices of LRSM}
	\label{sec2a}
	As mentioned earlier, texture zeros in lepton mass matrices increase the predictive power of the model due to a decrease in the number of free parameters \cite{Ludl:2014axa,Xing:2002ta,Singh:2016qcf,Ahuja:2017nrf,Meloni:2014yea,Fritzsch:2011qv,Alcaide:2018vni,Zhou:2015qua,Bora:2016ygl,Borgohain:2018lro}. Since the light neutrino mass comes from a combination of type I seesaw term $M_D{M_{RR}}^{-1}{M_D}^T$ and a type II seesaw term $M_{LL} \propto M_{RR}$, the requirement of having allowed number of zeros in the light neutrino mass matrix can constrain the texture zeros in $M_D, M_{RR}$ in an interesting way. Although $M_{\nu}$ can have at most six zeros (only 6 out of 15 allowed), we can have more texture zero possibilities in $M_D, M_{RR}$. Since $M_D$ is not necessarily Hermitian, we can have nine independent elements so that $n$ texture zeros can have $^9C_n$ possibilities. On the other hand, $M_{RR}$,  being complex symmetric can have six independent elements will have $^6C_n$ possibilities for $n$ texture zeros. While finding texture zeros in $M_{RR}$ we, however,  make sure that the determinant is non-zero so that the type I seesaw formula can be applied.
	We classify these texture zero possibilities as follows.
	\begin{itemize}
		\item The different classes of 4-0 texture $\rm M_{RR}$ with non zero determinant are:
		\begin{equation}\label{eq22}
		\rm M_{RR}=\left[\begin{array}{ccc}
		0 & w & 0\\
		w& 0& 0\\
		0& 0& x
		\end{array}\right],\rm M_{RR}=\left[\begin{array}{ccc}
		0 & 0 &w \\
		0 & x& 0\\
		w& 0& 0
		\end{array}\right],\rm M_{RR}=\left[\begin{array}{ccc}
		w & 0 &0 \\
		0 & 0& x\\
		0& x& 0
		\end{array}\right]
		\end{equation}
		\item The different classes of 3-0 texture $\rm M_{RR}$ with  non zero determinant are:
		\begin{equation}\label{eq23}
		\rm M_{RR}=\left[\begin{array}{ccc}
		0 & 0 & w\\
		0& x& y\\
		w& y& 0
		\end{array}\right],\rm M_{RR}=\left[\begin{array}{ccc}
		0 & w &x \\
		w & 0& y\\
		x& y& 0
		\end{array}\right],\rm M_{RR}=\left[\begin{array}{ccc}
		0 & w &0 \\
		w & x& 0\\
		0& 0& y
		\end{array}\right]\\
		\end{equation}
		
		\begin{equation}\label{eq24}
		\rm M_{RR}=\left[\begin{array}{ccc}
		0 & 0 & w\\
		0& x& 0\\
		w& 0& y
		\end{array}\right],\rm M_{RR}=\left[\begin{array}{ccc}
		0 & w &0 \\
		w & 0& x\\
		0& x&y
		\end{array}\right],\rm M_{RR}=\left[\begin{array}{ccc}
		w& 0 &0 \\
		0 & 0& x\\
		0& x& y
		\end{array}\right]\\
		\end{equation}
		
		\begin{equation}\label{eq25}
		\rm M_{RR}=\left[\begin{array}{ccc}
		w& 0 &0\\
		0& x& y\\
		0& y& 0
		\end{array}\right],\rm M_{RR}=\left[\begin{array}{ccc}
		w & 0 &0 \\
		0 & x& 0\\
		0& 0& y
		\end{array}\right],\rm M_{RR}=\left[\begin{array}{ccc}
		w & x &0 \\
		x & 0& 0\\
		0& 0& z
		\end{array}\right]\\
		\end{equation}
		
		\begin{equation}\label{eq26}
		\rm M_{RR}=\left[\begin{array}{ccc}
		w & x & 0\\
		x& 0& y\\
		0& y& 0
		\end{array}\right],\rm M_{RR}=\left[\begin{array}{ccc}
		0 & w &0 \\
		w & 0& x\\
		0& x& y
		\end{array}\right],\rm M_{RR}=\left[\begin{array}{ccc}
		0 & w &0 \\
		w & x& 0\\
		0& 0& y
		\end{array}\right]\\
		\end{equation}
		
		\begin{equation}\label{eq27}
		\rm M_{RR}=\left[\begin{array}{ccc}
		w & x & 0\\
		x& 0& y\\
		0& y& 0
		\end{array}\right]\\
		\end{equation}	
		
		\item The different classes of 2-0 texture $\rm M_{RR}$ with  non zero determinant are:	
		
		\begin{equation}\label{eq28}
		\rm M_{RR}=\left[\begin{array}{ccc}
		0 & 0 & w\\
		0& x& y\\
		w& y& z
		\end{array}\right],\rm M_{RR}=\left[\begin{array}{ccc}
		0 & x &0 \\
		x & y& z\\
		0&z& u
		\end{array}\right],\rm M_{RR}=\left[\begin{array}{ccc}
		w & x &0 \\
		x & 0& y\\
		0& y& z
		\end{array}\right]\\
		\end{equation}
		
		\begin{equation}\label{eq29}
		\rm M_{RR}=\left[\begin{array}{ccc}
		w & 0 & x\\
		0& y& z\\
		x& z& u
		\end{array}\right],\rm M_{RR}=\left[\begin{array}{ccc}
		w & 0 &x \\
		0 & 0&y\\
		x& y&z
		\end{array}\right],\rm M_{RR}=\left[\begin{array}{ccc}
		w& x &0 \\
		x & y& z\\
		0& z& u
		\end{array}\right]\\
		\end{equation}
		
		\begin{equation}\label{eq30}
		\rm M_{RR}=\left[\begin{array}{ccc}
		w& x &y\\
		x& 0& z\\
		y& z& 0
		\end{array}\right],\rm M_{RR}=\left[\begin{array}{ccc}
		w & x &y \\
		x & 0& 0\\
		0& 0& z
		\end{array}\right],\rm M_{RR}=\left[\begin{array}{ccc}
		w & x &y \\
		x & z& 0\\
		y& 0& 0
		\end{array}\right]\\
		\end{equation}
		
		\begin{equation}\label{eq31}
		\rm M_{RR}=\left[\begin{array}{ccc}
		0 & w & x\\
		w& 0& y\\
		x& y& z
		\end{array}\right],\rm M_{RR}=\left[\begin{array}{ccc}
		0 & w &x \\
		w & y& z\\
		x& z& 0
		\end{array}\right],\rm M_{RR}=\left[\begin{array}{ccc}
		0 & w &x \\
		w & y& 0\\
		x& 0& z
		\end{array}\right]\\
		\end{equation}
		
		\begin{equation}\label{eq32}
		\rm M_{RR}=\left[\begin{array}{ccc}
		w & 0 & 0\\
		0& x& y\\
		0& y&z
		\end{array}\right],\rm M_{RR}=\left[\begin{array}{ccc}
		w & 0 &x \\
		0 & y& 0\\
		x& 0& z
		\end{array}\right],\rm M_{RR}=\left[\begin{array}{ccc}
		w & x &0 \\
		x & y& 0\\
		0& 0& z
		\end{array}\right]\\
		\end{equation}	
		
		\item The different classes of 1-0 texture $\rm M_{RR}$ with  non zero determinant are:	
		
		\begin{equation}\label{eq33}
		\rm M_{RR}=\left[\begin{array}{ccc}
		0 & w & x\\
		w& y& z\\
		x& z& u
		\end{array}\right],\rm M_{RR}=\left[\begin{array}{ccc}
		w & 0 &x \\
		0 & y& z\\
		x&z& u
		\end{array}\right],\rm M_{RR}=\left[\begin{array}{ccc}
		w & x &0 \\
		x & 0& y\\
		0& y& z
		\end{array}\right]\\
		\end{equation}
		
		\begin{equation}\label{eq34}
		\rm M_{RR}=\left[\begin{array}{ccc}
		w & x & y\\
		x& 0& z\\
		y& z& u
		\end{array}\right],\rm M_{RR}=\left[\begin{array}{ccc}
		w & x &y \\
		x & z&0\\
		y& 0&u
		\end{array}\right],\rm M_{RR}=\left[\begin{array}{ccc}
		w& x &y \\
		x & z& u\\
		y& u& 0
		\end{array}\right]\\
		\end{equation}

	\end{itemize}
	
	\begin{table}[h!]
		\centering
		\begin{tabular}{|c|| c|c| c| c|c ||  }
			\hline
			$M_{RR}$ and $M_D$ textures&Total textures&1-0(A)&	2-0(A) & No-0(A)&Total (A) \\ \hline\hline\hline
			5-0 $M_D$, 4-0 $M_{RR}$& 378&62 &109	&18&189 \\ \hline\hline
			5-0 $M_D$, 3-0 $M_{RR}$& 1638	&628&23	&481&1132\\ \hline\hline
			5-0 $M_D$, 2-0 $M_{RR}$&1890	&553&73	&1155&1781\\ \hline\hline
			4-0 $M_D$, 4-0 $M_{RR}$	&378&161&76	&70&307\\ \hline\hline
			4-0 $M_D$, 3-0 $M_{RR}$	&1638&504&114&928&1546\\ \hline\hline
			4-0 $M_D$, 2-0 $M_{RR}$	&1890&277&34	&1534&1845\\ \hline\hline
			4-0 $M_D$, 1-0 $M_{RR}$	&756&40&	&716&756\\ \hline\hline
			3-0 $M_D$, 4-0 $M_{RR}$	&252&78&	&133&211\\ \hline\hline
			3-0 $M_D$, 3-0 $M_{RR}$	&1092&168&19	&896&1083\\ \hline\hline
			3-0 $M_D$, 2-0 $M_{RR}$	&1260&68&6	&1179&1253\\ \hline\hline
			3-0 $M_D$, 1-0 $M_{RR}$	&504&9&	&495&504\\ \hline\hline	
			2-0 $M_D$, 4-0 $M_{RR}$	&108&12&	&96&108\\ \hline\hline
			2-0 $M_D$, 3-0 $M_{RR}$	&468&15&	&453&468\\ \hline\hline
			2-0 $M_D$, 2-0 $M_{RR}$	&540&4&	&536&540\\ \hline\hline
			2-0 $M_D$, 1-0 $M_{RR}$	&&&	&216&216\\ \hline\hline
			1-0 $M_D$, 4-0 $M_{RR}$	&27&&	&27&27\\ \hline\hline
			1-0 $M_D$, 3-0 $M_{RR}$	&117&&	&117&117\\ \hline\hline
			1-0 $M_D$, 2-0 $M_{RR}$	&135&&	&135&135\\ \hline\hline
			1-0 $M_D$, 1-0 $M_{RR}$&54&&	&54&54\\ \hline\hline
		\end{tabular}
		\caption{Different number of allowed (A) texture zero neutrino mass for different textures of $M_D$ and $M_{RR}$. The blank boxes mean no possibilities.} \label{1}
	\end{table}
	The different number of allowed texture structures obtained for the various combinations of 5-0, 4-0, 3-0, 2-0 and 1-0 $M_D$ with 4-0, 3-0, 2-0, 1-0 $M_{RR}$ are shown in tabular form in table \ref{1}. However, for detailed numerical analysis, we will consider the right-handed Majorana mass matrix with the highest number of zeros, i.e 4-0 texture $M_{RR}$ as given by equation \ref{eq22}. Similarly, we will consider $M_D$ with 5 zeros (maximum) which can phenomenologically provide the allowed zero textures in the light neutrino mass matrix.
	\begin{table}[h!]
		\centering
		\begin{tabular}{|c|| c|c| c| c| c| c| c| }
			\hline
			$M_{RR}$&1-0(A)&	2-0(A) & No-0(A)   & 2-0(NA)& 3-0(NA)&4-0(NA)&Total $M_D$\\ \hline
			1&20 &27	&6 &48  &23 &2&126\\ \hline\hline\hline
			2&20 &27  	&6 &51 &20  &2&126\\ \hline\hline\hline
			3&22 &55    &6 &21 &20 &2&126\\ \hline	
		\end{tabular}
		\caption{Number of different textures obtained for 5-0 $M_D$, 4-0 $M_{RR}$ (with rank 3). A and NA in brackets represent allowed and not allowed cases.} \label{2}
	\end{table}
	Furthermore, from table \ref{2}, we will take into consideration only the allowed cases of two texture zero structures of light neutrino mass matrix. Out of a total of $^6C_2$ i.e., 15 two texture zeros of $\nu$ mass matrix, 6 are totally allowed by neutrino and cosmology data. It should be noted that these conclusions hold for diagonal charged lepton basis which we also adopt in our analysis. These allowed two zero texture light neutrino mass matrices are given as
	
	\begin{equation}\label{eq35}
	A1=\left[\begin{array}{ccc}
	0 & 0 & \times\\
	0& \times& \times\\
	\times& \times& \times 
	\end{array}\right],A2=\left[\begin{array}{ccc}
	0 & \times &0 \\
	\times & \times& \times\\
	0& \times& \times
	\end{array}\right]
	\end{equation}

	\begin{equation}\label{eq36}
	B1=\left[\begin{array}{ccc}
	\times & \times &0\\
	\times& 0& \times\\
	0 & \times&\times
	\end{array}\right],B2=\left[\begin{array}{ccc}
	\times & 0&\times \\
	0 & \times& \times\\
	0& \times& 0
	\end{array}\right], B3=\left[\begin{array}{ccc}
	\times & 0& \times \\
	0 & 0 & \times\\
	\times & \times & \times
	\end{array}\right], B4=\left[\begin{array}{ccc}
	\times & \times& 0 \\
	\times & \times & \times\\
	0 & \times & 0
	\end{array}\right]
	\end{equation}
	where $\times$ denotes any non-zero entry. Since we have only three possible $M_{RR}$ structures with non zero determinants (as given in equations \ref{eq22}), the possibilities of obtaining the allowed two zero texture neutrino mass matrix for a particular texture of $M_{RR}$ are also limited. The allowed two zero textures obtained for the three different $M_{RR}$ textures are (A2, B1), (A1, B2) and (B1, B2, B3, B4) respectively for $M_D$ with five zeros. Herein we have picked up these combinations of $M_D$ and $M_{RR}$ which lead to the allowed class of two zero texture neutrino mass in the framework of minimal LRSM.
	\begin{itemize}
		
		\item\textbf{For the class A1 ($M_{ee}=0, M_{e\mu}=0$)}

		\begin{equation}\label{eq37}
		\rm M_{RR}=\left[\begin{array}{ccc}
		0 & 0 &w \\
		0 & x& 0\\
		w& 0& 0
		\end{array}\right],\rm M_D=\left[\begin{array}{ccc}
		0 & 0 &a_3 \\
		0 & b_2& 0\\
		0& c_2& c_3
		\end{array}\right]
		\end{equation}
		\item\textbf{For the class A2 ($M_{ee}=0, M_{e\tau}=0$)}

		\begin{equation}\label{eq38}
		\rm M_{RR}=\left[\begin{array}{ccc}
		0 & w &0 \\
		w & 0& 0\\
		0& 0& x
		\end{array}\right],\rm M_D=\left[\begin{array}{ccc}
		0 & 0 &0 \\
		0 & b_3& 0\\
		c_1& c_2& c_3
		\end{array}\right]
		\end{equation}
		\item\textbf{For the class B1 ($M_{e\tau}=0, M_{\mu\mu}=0 $)}
		\begin{equation}\label{eq39a}
		\rm M_{RR}=\left[\begin{array}{ccc}
		w & 0 &0 \\
		0 & 0& x\\
		0& x& 0
		\end{array}\right],\rm M_D=\left[\begin{array}{ccc}
		0 & a_2 &a_3 \\
		0 & b_2& 0\\
		c1& 0& 0
		\end{array}\right]
		\end{equation}
		\begin{equation}\label{eq39b}
		\rm M_{RR}=\left[\begin{array}{ccc}
		0 & w &0 \\
		w & 0& 0\\
		0& 0& x
		\end{array}\right],\rm M_D=\left[\begin{array}{ccc}
		a_1 & 0 &a_3 \\
		0 & b_2& 0\\
		c_1& 0& 0
		\end{array}\right]
		\end{equation}
		\item\textbf{For the class B2 ($M_{e\mu}=0, M_{\tau\tau}=0$)}
		\begin{equation}\label{eq40a}
		\rm M_{RR}=\left[\begin{array}{ccc}
		w & 0 &0 \\
		0 & 0& x\\
		0& x& 0
		\end{array}\right],\rm M_D=\left[\begin{array}{ccc}
		0 & 0 &a_3 \\
		b_1 & 0& b_3\\
		0& c_2& 0
		\end{array}\right]
		\end{equation}
		\begin{equation}\label{eq40b}
		\rm M_{RR}=\left[\begin{array}{ccc}
		0 & 0 &w \\
		0 & x& 0\\
		w& 0& 0
		\end{array}\right],\rm M_D=\left[\begin{array}{ccc}
		a_1 & a_2 &0 \\
		b_1 & 0& 0\\
		0& 0& c_3
		\end{array}\right]
		\end{equation}
		\item\textbf{For the class B3 ($M_{e\mu}=0, M_{\mu\mu}=0$)}
		
		\begin{equation}\label{eq41}
		\rm M_{RR}=\left[\begin{array}{ccc}
		w & 0 &0 \\
		0 & 0& x\\
		0& x& 0
		\end{array}\right],\rm M_D=\left[\begin{array}{ccc}
		0 & a_2 &a_3 \\
		0 & 0& 0\\
		c_1& 0& c_3
		\end{array}\right]
		\end{equation}
		
		\item\textbf{For the class B4 ($M_{\mu\mu}=0, M_{\tau\tau}=0)$}
		
		\begin{equation}\label{eq42}
		\rm M_{RR}=\left[\begin{array}{ccc}
		w & 0 &0 \\
		0 & 0& x\\
		0& x& 0
		\end{array}\right],\rm M_D=\left[\begin{array}{ccc}
		0 & a_2 &0 \\
		b_1 & b_2& b_3\\
		0& 0& 0
		\end{array}\right]
		\end{equation}
		
	\end{itemize}
	Although our study is motivated from a phenomenological point of view, it is worth mentioning that the texture zeros in fermion mass matrices can have dynamical origin from flavour symmetries. See, for example, the scenarios proposed in \cite{Gu:2008yj, Deppisch:2012vj, CarcamoHernandez:2018hst, Lamprea:2016egz, delaVega:2018cnx, Cebola:2015dwa, Berger:2000zj, Grimus:2004hf, Dev:2011jc, Araki:2012ip, Felipe:2014vka} where discrete as well as continuous symmetries were considered to explain the texture zeros. In particular, the recent work \cite{CarcamoHernandez:2018hst} considered a different version of LRSM where charged fermions receive masses from a universal seesaw mechanism while neutrinos acquire masses at the radiative level. A non-abelian discrete flavour symmetry based on the $\triangle (27)$ group was incorporated, leading to predictive textures of different fermion mass matrices. We leave such a flavour symmetric explanation of the textures considered here for future works.
	
	Phenomenological implications of two texture zero $M_\nu$ on low energy phenomena like NDBD and CLFV have been analyzed in one of our earlier work \cite{Borgohain:2018lro}. However, in that case, the authors have considered the two zero texture mass matrix to be favouring a tri-maximal mixing pattern.  Besides, all the contributions to NDBD that could arise in the framework of LRSM were not taken into consideration. Here we generalize this to consider maximum allowed texture zeros that is 5-0 $M_D$ and 4-0 $M_{RR}$ giving rise to the allowed two zero texture neutrino mass matrix and then study the implications of these $M_D$ and $M_{RR}$ for NDBD, considering all the possible contributions that could arise in LRSM and also study for lepton flavour violating processes like  $\mu\rightarrow e\gamma$ and $\mu\rightarrow 3e$.
	\section{Neutrinoless Double Beta Decay in LRSM}
	{\label{sec3}}
	Neutrinoless double beta decay is a process where a nucleus emits two electrons thereby changing its atomic number by two units
	$$ (A, Z) \rightarrow (A, Z+2) + 2e^- $$
	with no neutrinos in the final state. Such a process violates lepton number by two units and hence is a probe of Majorana neutrinos, which are predicted by generic seesaw models of neutrino masses. For a review and recent status of NDBD, please refer to \cite{Rodejohann:2011mu, Cardani:2018lje, Dolinski:2019nrj}. Apart from probing the intrinsic nature of light neutrinos, NDBD can also be used to discriminate between neutrino mass hierarchies: normal versus inverted. From the measurement of NDBD half-life combined with sufficient information about the phase space factors (PSF) and associated nuclear matrix element (NME), one can set constraints on the absolute neutrino mass scales. If light neutrinos are Majorana, we can get a sizeable contribution to NDBD especially when the ordering is of inverted type. There have been several works where BSM contributions to NDBD have been calculated. For example, see \cite{Mohapatra:1986su,Babu:1995vh,Hirsch:1995vr,Hirsch:1996ye,Deppisch:2012nb,Schechter:1981bd,Ge:2015yqa,Allanach:2009xx} and references therein. In the LRSM scenario, it has been widely studied in several earlier works including \cite{Awasthi:2013ff,Patra:2012ur,Chakrabortty:2012mh,Tello:2010am,Awasthi:2015ota,Huang:2013kma,Borah:2016iqd,Borah:2015ufa,Hirsch:1996qw,Bambhaniya:2015ipg,Dev:2014xea,Barry:2013xxa,Borgohain:2017akh}. Owing to the presence of many new heavy particles in LRSM, sizeable new contributions of NDBD decay amplitudes arises which may be dominant over the standard mechanism mediated by light neutrinos. Out of the different NDBD experiments, KamLAND-Zen \cite{KamLAND-Zen:2016pfg} has reported a strong lower limit on the half-life from searches on $ ^136 Xe$ as $ \rm T_{1/2}^{0\nu}>1.07\times 10^{26}$ year at $ 90\%$ C. L. This can be translated to an upper limit of effective Majorana mass in the range $ (0.061-0.165)$ eV where the uncertainty arises due to the NME. 
	

	\begin{figure}[h!]
		\centering
		\includegraphics[width=0.39\textwidth,height=4.82cm]{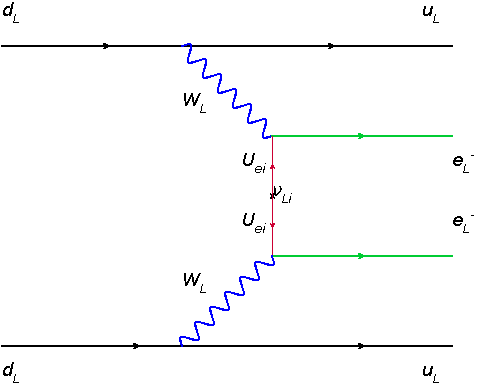}
		\includegraphics[width=0.39\textwidth,height=4.82cm]{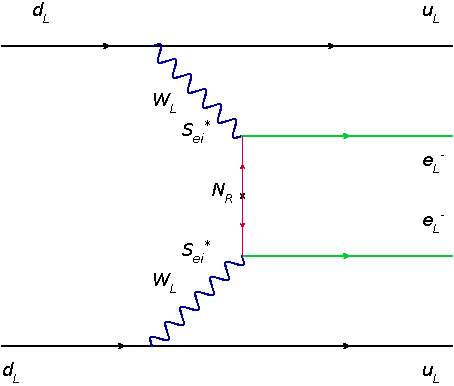}

		\caption{ Feynman diagrams corresponding to neutrinoless double beta decay due to $\nu-W_L-W_L$, $N-W_L-W_L$contributions.  } \label{figa}
	\end{figure}
	\begin{figure}[h!]
		\centering
		
		\includegraphics[width=0.39\textwidth,height=4.82cm]{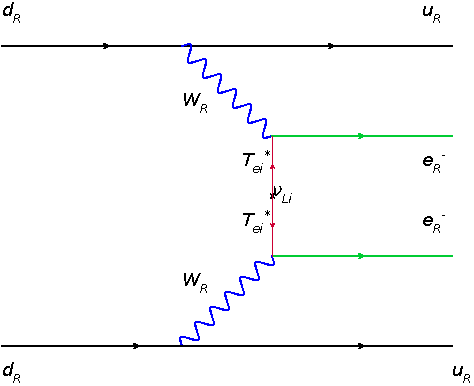}
		\includegraphics[width=0.39\textwidth,height=4.82cm]{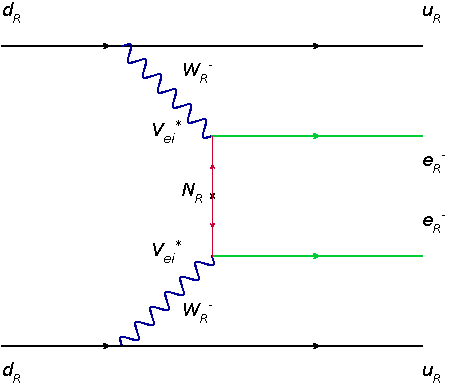}
		
		\caption{ Feynman diagrams corresponding to neutrinoless double beta decay due to $\nu-W_R-W_R$, , $N-W_R-W_R$  contributions. } \label{figb}
	\end{figure}
	\begin{figure}[h!]
		\centering
		
		\includegraphics[width=0.39\textwidth,height=4.82cm]{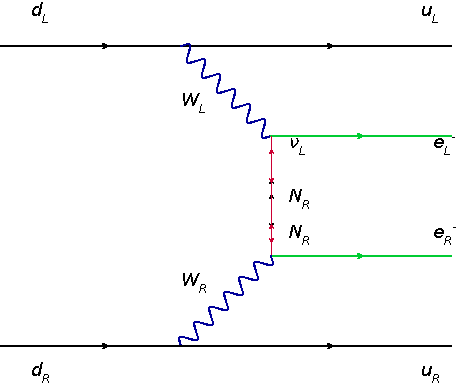}
		\includegraphics[width=0.39\textwidth,height=4.82cm]{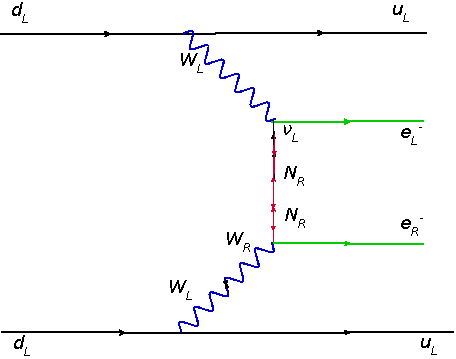}

		\caption{ Feynman diagrams corresponding to neutrinoless double beta decay due to  $N-W_L-W_R$ mediation with heavy-light neutrino exchange and $W_L-W_R$ mixing ($\lambda$ and $\eta$ contributions).  } \label{figc}
	\end{figure}
	
	\begin{figure}[h!]
		\centering
		
		\includegraphics[width=0.39\textwidth,height=4.82cm]{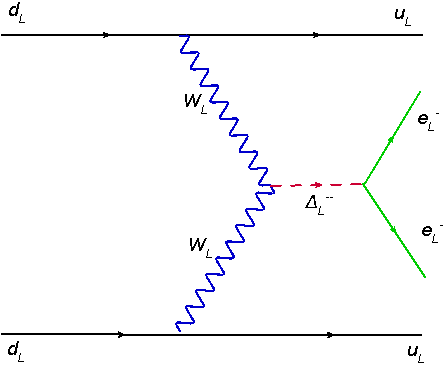}
		\includegraphics[width=0.39\textwidth,height=4.82cm]{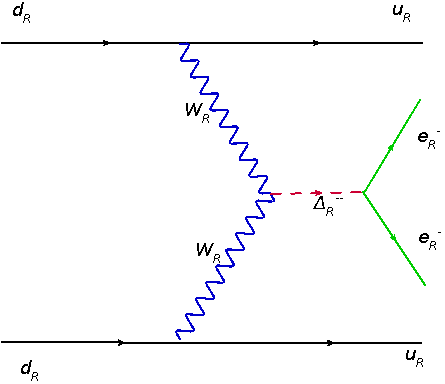}
		
		\caption{ Feynman diagrams corresponding to neutrinoless double beta decay due to  $\Delta_L-W_L$ $\Delta_R-W_R$ contributions.  } \label{figd}
	\end{figure}
	We show all the contributions to NDBD in minimal LRSM in terms of corresponding Feynman diagrams in figure \ref{figa}, \ref{figb}, \ref{figc}, \ref{figd}. We now list their respective contributions below one by one following the notations of \cite{Barry:2013xxa}.
	\begin{itemize}
		
		\item 	When light and heavy neutrinos are the source of NDBD mediated by purely left handed (LH) currents  ($W_L-W_L$ ) as shown in  figure \ref{figa}, the corresponding amplitudes are given by,
		\begin{equation}\label{eqa1}
		\rm{A_\nu}^{LL} \propto G_{F}^2\sum_i \frac{{U_{{e_i}}}^2m_i}{p^2},\rm{A_N}^{LL} \propto G_{F}^2\sum_i \frac{{S_{{e_i}}}^2M_i}{p^2}.  
		\end{equation}
		where, $\left|p\right|\sim$ 100 MeV is the typical momentum transfer at the leptonic vertex, U and S represent the mixing matrices as given in equations \ref{eq17} and \ref{eq18}, $m_i$ and $M_i$ are the masses for the three generations of light and heavy Majorana neutrinos respectively.
		\item The right-handed current mediated by $W_R$ can contribute to NDBD through the exchange of the light as well as heavy neutrino N (as shown in figure \ref{figb}). The corresponding amplitudes are given by,
		\begin{equation}\label{eqa2}
		\rm{A_\nu}^{RR} \propto G_{F}^2\sum_i \left(\frac{M_{W_L}}{M_{W_R}}\right)^4\left(\frac{g_R}{g_L}\right)^4\frac{{T_{e_i}^*}^2m_i}{p^2},
		\end{equation}
		\begin{equation}\label{eqa3}
		\rm{A_N}^{RR} \propto G_{F}^2\sum_i \left(\frac{M_{W_L}}{M_{W_R}}\right)^4\left(\frac{g_R}{g_L}\right)^4\frac{{V_{e_i}^*}^2M_i}{p^2},
		\end{equation}
		where, $M_{W_L}$ and $M_{W_R}$ are the mass of the LH and RH gauge bosons respectively.
		\item  Significant Contribution can also arise due to the mixed helicity diagrams, mediated by both  $W_L$ and $W_R$ ($\lambda$ contribution) and from diagrams mediated by $W_L-W_R$ mixing ($\eta$ contribution), the amplitudes of which are given as,
		\begin{equation}\label{eqa4}
		\rm{A_\lambda} \propto G_{F}^2\sum_i \left(\frac{M_{W_L}}{M_{W_R}}\right)^2\frac{ U_{e_i}T_{e_i}^*}{p}, 	\rm{A_\eta} \propto G_{F}^2\sum_i tan \xi\frac{ U_{e_i}T_{e_i}^*}{p},
		\end{equation}
		where $\xi$ is the L-R gauge boson mixing parameter as described earlier.
		\item Further, there is also the scalar triplet $(\Delta_{L,R})$ contributions to NDBD by the mediations of $W_L$ and $W_R$ gauge bosons respectively, the amplitude of which depends upon the masses of these gauge bosons and given by,
		\begin{equation}\label{eqa5}
		\rm{A_{\Delta_L}}\propto G_{F}^2 \frac{{\left({M_\nu}^{II}\right)}_{ee}}{{M_{\Delta_L}^{++}}^2}, 	
		\rm A_{\Delta_R}\propto G_{F}^2 \left(\frac{M_{W_L}}{M_{W_R}}\right)^4\frac{{V_{e_i}}^2 M_i}{{M_{\Delta_R}^{++}}^2} 
		\end{equation}	
		where the contribution from left triplet scalar is negligible due to smallness of $v_L$ as well as the smallness of light neutrino mass contribution coming from type II seesaw.
	\end{itemize}
	The particle physics parameters governing NDBD for the different contributions (ignoring the left triplet Higgs contribution) in LRSM we have considered are given by,
	\begin{equation}\label{eq43}
	\left|\eta_{\nu}\right|=\frac{1}{m_e}\sum_i U_{e_i}^2 m_i
	\end{equation}
	\begin{equation}\label{eq44}
	\left|\eta_{N_R}^L\right|=m_p\sum_i \frac{S_{e_i}^2}{ m_i}
	\end{equation}
	
	\begin{equation}\label{eq45}
	\left|\eta_{{N_R}+{\Delta_R}}^R\right|=m_p\left(\frac{M_{W_L}}{M_{W_R}}\right)^4\left(\sum_i \frac{V_{e_i}^2} {M_i}+\sum_i\frac{V_{e_i}^2 M_i}{{M_{\Delta_R}}^2}\right)
	\end{equation}
	\begin{equation}\label{eq46}
	\left|\eta_\lambda\right|=\left(\frac{M_{W_L}}{M_{W_R}}\right)\left|\sum_i U_{e_i} T^*_{e_i}\right|.
	\end{equation}
	
	\begin{equation}\label{eq47}
	\left|\eta_\eta\right|= \tan \xi \left|\sum_i U_{ei} T^*_{ei}\right|.
	\end{equation}
	\par In the above equations, $ m_p$ and $ m_e$ are the mass of the proton and electron respectively. It is seen that the amplitudes of
	these processes are mostly dependent on the mixing between neutrinos, the mass of the heavy neutrinos, $\rm N_i$, the mass of the gauge bosons, $\rm {W_L}^-$ and $\rm {W_R}^-$,
	mass of doubly charged scalars triplet Higgs, $\rm \Delta_L$ and $\rm \Delta_R$  as well as their coupling to leptons, $\rm f_L$ and  $f_R$. The total analytic expression for the inverse half-life governing NDBD considering all the dominant contributions that could arise in LRSM is given by,
	\begin{equation}\label{eq48}
	\left[{T_{\frac{1}{2}}}^{0\nu}\right]^{-1}=G^{0\nu}(Q,Z)\left({\left|M^{0\nu}_\nu\eta_\nu+M^{0\nu}_N\eta_{N_R}^L \right|}^2+{\left|M^{0\nu}_N\eta_{N_R}^R+M^{0\nu}_N\eta_{\Delta_R}\right|}^2+{\left|M^{0\nu}_\lambda\eta_\lambda+M^{0\nu}_\eta\eta_\eta\right|}^2\right),
	\end{equation}
	In the above expression, $G^{0\nu}(Q,Z)$ represents the phase space factor and $M^{0\nu}$ is the nuclear matrix element which have different values for different contributions which is shown in tabular form in table \ref{3} \cite{Dev:2014xea}.

	\begin{table}[h!]
		\centering
		\begin{tabular}{|c|| c|c| c| c| c| c| c| }
			\hline
			Isotope	&	$G^{0\nu}(Q,Z)(yr^{-1})$&$M^{0\nu}_\nu$&$M^{0\nu}_N$& $M^{0\nu}_\lambda$  &  $M^{0\nu}_\eta$  \\ \hline
			$76_{Ge}$	&	5.77$\times 10^{-15}$&2.58-6.64 &233-412	&1.75-3.76 &235-637  \\ \hline\hline\hline
			$136_{Xe}$	&	3.56$\times 10^{-14}$&1.57-3.85 &164-172  	&1.92-2.49 &370-419 \\ \hline\hline\hline		
		\end{tabular}
		\caption{The different values of PSF and NME for different nuclei used in NDBD experiments.} \label{3}
	\end{table}

	\section{Charged Lepton Flavour Violation in LRSM}
	\label{sec3a}
	Charged lepton flavour violation arises in the SM at one loop level and remains suppressed by the smallness of neutrino masses, much beyond the current and near future experimental sensitivities. Therefore, any experimental observation of such processes is definitely a sign of BSM physics, like the one we are studying here. For a review of CLFV in SM and beyond, please refer to \cite{Lindner:2016bgg}.
	Though usual light neutrino contribution to CLFV is negligible, presence of heavy neutrinos in BSM frameworks can give rise to observable CLFV \cite{Leontaris:1985qc,Swartz:1989qz,Cirigliano:2004mv,Cirigliano:2004tc,Bajc:2009ft,Barry:2013xxa, Bernstein:2013hba,Borah:2016iqd, Borgohain:2017inp, Bambhaniya:2015ipg,FileviezPerez:2017zwm}. In LRSM, sizeable CLFV occurs dominantly due to the contributions arising from the additional scalars and the heavy neutrinos. Among the various processes that violate lepton flavour, the most relevant ones are the rare leptonic decay modes of the muon, notably, $\left(\mu\rightarrow e\gamma\right)$ and $ \left(\rm \mu\rightarrow 3e \right) $. The best upper limit for the branching ratio (BR) of these processes are provided by MEG collaboration \cite{Baldini:2013ke} and SINDRUM experiment \cite{Bellgardt:1987du} which provide the corresponding upper limit as $ {\rm BR}\left(\mu\rightarrow e\gamma\right) <4.2\times 10^{-13}$ and  ${\rm BR}\left(\mu\rightarrow 3e\right)<1.0\times 10^{-12}$ respectively.
	
	Adopting the notations of \cite{Barry:2013xxa, Borah:2016iqd} the branching ratio of the process $\mu\rightarrow 3e$ mediated by doubly charged scalars can be written as
	\begin{equation}\label{eq49}
	\rm BR \left(\mu\rightarrow 3e\right)=\frac{1}{2}{\left|h'_{\mu e}{h'_{ee}}^{*}\right|}^{2}\left(\frac{{M_{W_L}}^4}{{M_{\Delta_L}^{++}}^4}+\frac{{M_{W_R}}^4}{{M_{\Delta_R}^{++}}^4}\right),
	\end{equation}
	where $h'_{ij}$ describes the respective lepton-scalar couplings given by, 
	\begin{equation}\label{eq50}
	\rm h'_{ij}=\sum_{n=1}^{3}V_{in}V_{jn}\left(\frac{M_n}{M_{W_R}}\right), i,  j=e,\mu,\tau.
	\end{equation}
	with $V$ being one of the lepton mixing matrices given in \eqref{eq17}.
	
	The branching ratio for the CLFV process $\mu\rightarrow e\gamma $ is given by (as explained in \cite{Barry:2013xxa}),
	\begin{equation}\label{eq51}
	\rm BR\left(\mu\rightarrow e\gamma\right)= \frac{3\alpha_{em}}{2\Pi}\left(\left|G_{L}^\gamma\right|^2+\left|G_{R}^\gamma\right|^2\right) ,
	\end{equation}\\
	where, $ \alpha_{em}$ is the fine structure constant defined as $ \alpha_{em}=\frac{e^2}{4\Pi}$, $G_{L}^\gamma$ and $G_{R}^\gamma$ are the form factors given by,
	
	\begin{equation}\label{eq52}
	G_{L}^\gamma=\sum_{i=1}^{3}\left({S_{\mu i}}^*S_{e i}G_{1}^\gamma(x_i)-V_{\mu i}S_{e i}\xi e^{i\zeta}G_{2}^\gamma(x_i)\frac{M_i}{m_{\mu}}+V_{\mu i}{V_{e i}}^*y_i\left[\frac{2}{3}\frac{{M_{W_L}^2}}{M_{{\Delta_L}^{++}}^2}+\frac{1}{12}\frac{{M_{W_L}^2}}{M_{{\Delta_L}^{+}}^2}\right]\right)
	\end{equation}
	\begin{equation}\label{eq53}
	G_{R}^\gamma=\sum_{i=1}^{3}\left(V_{\mu i}{V_{e i}}^*{\left|\xi\right|}^2G_{1}^\gamma(x_i)-{S_{\mu i}}^*{V_{e i}}^*\xi e^{-i\zeta}G_{2}^\gamma(x_i)\frac{M_i}{m_{\mu}}+V_{\mu i}{V_{e i}}^*\left[\frac{{M_{W_L}^2}}{M_{{W_R}}^2}G_{1}^\gamma(y_i)+\frac{2 y_i}{3}\frac{{M_{W_L}^2}}{M_{{\Delta_R}^{++}}^2}\right]\right).
	\end{equation}
	In the above equations, the terms $x_i=\left({\frac{M_i}{M_{W_L}}}\right)^2$ and $x_i=\left({\frac{M_i}{M_{W_R}}}\right)^2$, $M_{\Delta_{L,R}}$ are the masses of the left and right scalar triplets, $ \rm M_i(i=1,2,3) $ are the masses of the right-handed neutrinos.  V is the mixing matrix of the right-handed neutrinos given in \eqref{eq17}. $\zeta$ is the phase of the VEV $k_2$ which we consider to be negligible, whereas the left-right gauge boson mixing parameter, $\xi$ is also very small $\leqslant 10^{-6}$ in our case. S being the light-heavy neutrino mixing as defined in \ref{eq18}. Again the loop functions $G_{1,2}^\gamma(a)$ are defined as,
	\begin{equation}\label{eq54}
	G_{1}^\gamma(a)=-\frac{2a^3+5a^2-a}{4{(1-a)}^3}-\frac{3a^3}{2{(1-a)}^4}\rm ln a
	\end{equation}
	
	\begin{equation}\label{eq55}
	G_{2}^\gamma(a)=\frac{a^2-11a+4}{1{(1-a)}^2}-\frac{3a^2}{{(1-a)}^3}\rm ln a
	\end{equation}
	
	Recently the MEG collaboration has reported a new stringent upper bound on the decay rate of the process $\mu\rightarrow e\gamma$. The BR ratio for this LFV process as given by MEG is  $<4.2\times 10^{-13}$ at $ 90\%$ CL \cite{Baldini:2013ke}. While for the process $\mu\rightarrow 3e $ it is  $<1.0\times 10^{-12}$ as obtained  by the SINDRUM experiment \cite{Bellgardt:1987du}.

	\section{Numerical Analysis and Results}
\begin{figure}[h!]
		\centering
		\includegraphics[width=0.4\textwidth]{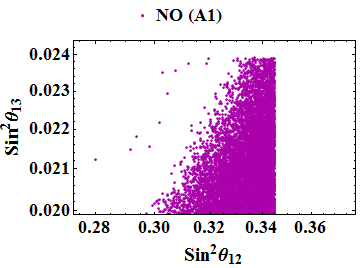}	
		\includegraphics[width=0.4\textwidth]{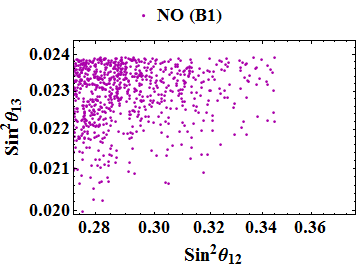} \\
		\includegraphics[width=0.4\textwidth]{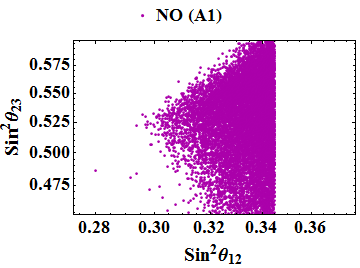}
		\includegraphics[width=0.4\textwidth]{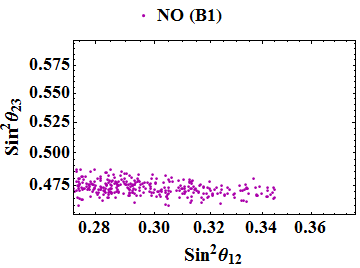}
		\caption{Correlation between light neutrino parameters in NO case.} \label{fig1}
		
	\end{figure}

\begin{figure}[h!]
		\centering
		\includegraphics[width=0.4\textwidth]{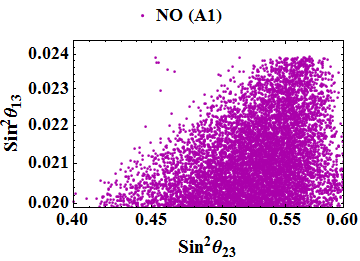}	
		\includegraphics[width=0.4\textwidth]{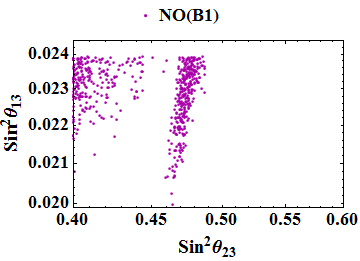} \\
		\includegraphics[width=0.4\textwidth]{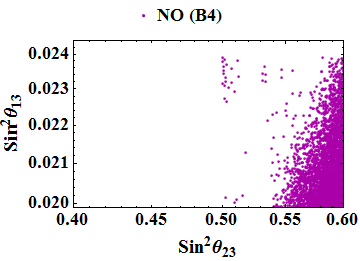}
		\includegraphics[width=0.4\textwidth]{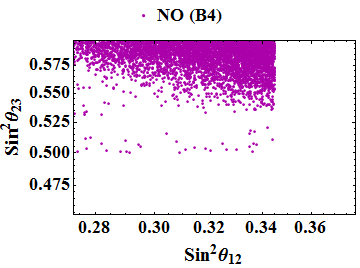}

		\caption{Correlation between neutrino parameters in NO case. } \label{fig2}	
	\end{figure}

\begin{figure}[h!]
		\centering
		\includegraphics[width=0.4\textwidth]{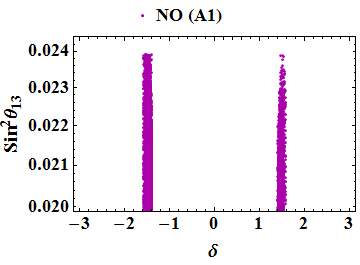}	
		\includegraphics[width=0.4\textwidth]{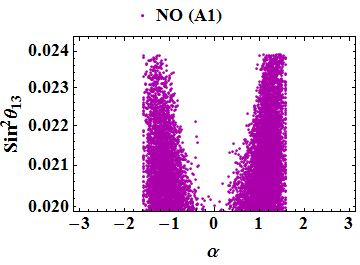} \\
		\includegraphics[width=0.4\textwidth]{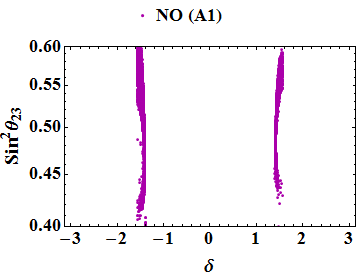}	
		\includegraphics[width=0.4\textwidth]{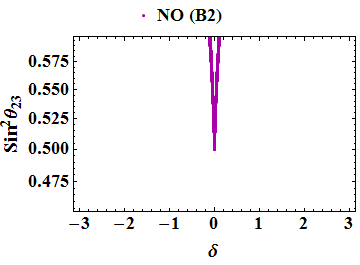}
		\caption{Correlation between neutrino parameters in NO case. } \label{fig3}	
	\end{figure}

{\label{sec4}}
	For our numerical analysis, we first parameterize the light neutrino mass matrix in terms of the Pontecorvo-Maki-Nakagawa-Sakata (PMNS) leptonic mixing matrix which is related to the diagonalizing 
	matrices of neutrino and charged lepton mass matrices $U_{\nu}, U_l$ respectively, as
	\begin{equation}
	U_{\text{PMNS}} = U^{\dagger}_l U_{\nu}
	\label{pmns0}
	\end{equation}
	The PMNS mixing matrix can be parametrized as
	\begin{equation}\label{eq59}
	U_{\text{PMNS}}=U_L=\left[\begin{array}{ccc}
	c_{12}c_{13}&s_{12}c_{13}&s_{13}e^{-i\delta}\\
	-c_{23}s_{12}-s_{23}s_{13}c_{12}e^{i\delta}&-c_{23}c_{12}-s_{23}s_{13}s_{12}e^{i\delta}&s_{23}c_{13}\\
	s_{23}s_{12}-c_{23}s_{13}c_{12}e^{i\delta}&-s_{23}c_{12}-c_{23}s_{13}s_{12}e^{i\delta}&c_{23}c_{13}
	\end{array}\right]P
	\end{equation}
	where $c_{ij} = \cos{\theta_{ij}}, \; s_{ij} = \sin{\theta_{ij}}$ and $\delta$ is the leptonic Dirac CP phase. The diagonal matrix $U_{\text{Maj}}=\text{diag}(1, e^{i\alpha}, e^{i\beta})$  contains the Majorana CP phases $\alpha, \beta$ which do not play any role in neutrino oscillations and hence are not constrained by neutrino data. In the diagonal charged lepton basis, considered in this work, we can write the light neutrino mass matrix as
	\begin{equation}\label{eq58}
	M_\nu= U_{\rm PMNS}{M_\nu}^{(\rm diag)} {U_{\rm PMNS}}^T
	\end{equation}
	where ${M_\nu}^{(\rm diag)} = {\rm diag}(m_1, m_2, m_3)$. We first implement the texture zero conditions on the light neutrino mass matrix and numerically solve the texture zero conditions to find the allowed parameter space. As pointed out earlier, there are six one zero texture possibilities whereas out of fifteen possible two zero textures, only six are compatible with neutrino and cosmology data which are labelled here as A1, A2, B1, B2, B3 and B4. Out of the nine parameters of the neutrino mass matrix, five are fixed by experimental measurements of the two mass-squared differences and three mixing angles. The remaining four parameters namely, $m_{\rm lightest} = m_1 ({\rm NO}) (m_3 ({\rm IO})), \delta, \alpha, \beta$ which are not measured yet, can be predicted by the texture zero conditions. This is possible in two zero texture cases particularly, because of two texture zero conditions which give rise to four real equations that can be solved simultaneously to find four unknown parameters. We vary the five known parameters randomly in the $3\sigma$ range using the recent global fit \cite{Esteban:2018azc}. Using the latest data, we found that out of the previously allowed six possible two zero textures, A2 for both NO and IO and A1 (IO) are disallowed. We consider the allowed ones for our analysis for NDBD and CLFV. For representative purpose, we show some correlations between light neutrino parameters coming out from the two zero texture conditions in figure \ref{fig1}, \ref{fig2}, \ref{fig3}. Similar correlation plots were obtained in earlier work \cite{Bora:2016ygl}.

	In minimal LRSM, the neutrino mass is given by equation \eqref{eq6} where the first and second terms represent the type II and type seesaw mass terms respectively. $\gamma$ is the dimensionless parameter that appears from the minimization of the scalar potential, defined before. We have fine-tuned the dimensionless parameter $\gamma=10^{-9}$ with a view to obtaining the neutrino mass of the order of sub eV. This is chosen particularly to keep the right-handed neutrino masses in the desired range. The right-handed neutrino mass matrix, defined earlier, is $M_{RR}=\sqrt{2}v_Rf_R = \frac{v_R}{v_L} {M_\nu}^{II} = \frac{v^2_R}{\gamma v^2_{\rm SM}} {M_\nu}^{II}$. The choice of $v_R$ for a few TeV $W_R$ mass, and type II seesaw term at sub-eV scale, the chosen value of $\gamma$ keeps the right-handed neutrino mass above 1 GeV. This is required to ensure that for the heavy neutrino mediated processes of NDBD, the masses of mediators remain above the typical momentum exchange of the process $\sim 100$ MeV. For heavy neutrino masses below this scale, the contribution to NDBD will be different, see for example \cite{Borah:2017ldt}. Recent ATLAS and CMS data enforce the $W_R$ boson to be heavier than about at least 3~TeV, the exact bound depending on the right-handed neutrino sector~\cite{Aaboud:2017efa, Aaboud:2017yvp, Sirunyan:2016iap, Khachatryan:2016jww, Sirunyan:2018mpc}. We consider it to be $M_{W_R}$ = 4.5 TeV, which satisfy the latest collider bounds \cite{Sirunyan:2018pom} for our chosen values of right-handed neutrino mass spectrum. All other relevant parameters of minimal LRSM which are used in the calculations are shown in table \ref{4}. It is worth noting that the chosen doubly charged scalar masses respect the latest bounds from collider experiment. On the basis of the results of the ATLAS searches for same-sign dileptonic new
	physics signals~\cite{Aaboud:2017qph}, there is a lower bound
	on the masses of the doubly-charged
	scalars $\Delta^{\pm \pm}_L$ and $\Delta^{\pm \pm}_R$. Assuming that the branching
	ratios into electronic and muonic final states are both equal to 50\%, the
	$SU(2)_L$ and $SU(2)_R$ doubly-charged Higgs-boson masses have to be larger than
	785~GeV and 675~GeV respectively. Our conservative lower bound on charged scalars from these triplets agree with all such experimental data.

	Having determined the light neutrino parameters which satisfy the two zero texture conditions, we then numerically determine the elements of $M_D, M_{RR}$ for the chosen textures. We then use the corresponding $M_D, M_{RR}$ as well as the light neutrino mass matrix for computing the relevant contributions to NDBD and CLFV. For NDBD mediated by the light Majorana neutrinos, the half-life of the decay process is given by,
	\begin{equation}\label{eq56}
	\frac{\Gamma_{0\nu\beta\beta}}{ln2}=\left(T_{1/2}^{O\nu}\right)^{-1}=G_O\nu\left|M^{0\nu}\right|^2\left|\frac{m_{\nu}^{\rm eff}}{m_e}\right|^2
	\end{equation}
	$\Gamma$ represents the decay width for $0\nu\beta\beta$ decay process.
	where $m_e$ is the electron mass and the terms $G^{0\nu}$ and $\left|M^{0\nu}\right|$ represents the phase space factor and the nuclear matrix elements respectively which holds different values as shown in table \ref{3}. The effective light neutrino mass is given by
	\begin{equation}\label{eq57}
	m_{\nu}^{\rm eff}=U_{Lei}^2m_i
	\end{equation}
	where, $U_{Lei}$ are the elements of the first row of the light neutrino mixing matrix. There are contributions coming from heavy right-handed neutrinos and right scalar Higgs triplets, both having exchange of $W_R$ bosons. The effective neutrino mass corresponding to these dominant contributions is given by,
	\begin{equation}\label{eq60}
	\rm {m_{N+\Delta_R}}^{eff}=p^2\frac{{M_{W_L}}^4}{{M_{W_R}}^4}\frac{{U_{Rei}}^*2}{M_i}+p^2\frac{{M_{W_L}}^4}{{M_{W_R}}^4}\frac{{{U_{Rei}}^2}M_i}{{M_{\Delta_R}}^2}.
	\end{equation}
	Here, $ \langle p^2 \rangle = m_e m_p \frac{M_N}{M_\nu}$ is the typical momentum exchange of the process, where $m_p$ and $ m_e$ are the mass of the proton and electron respectively and $ M_N$ is the nuclear matrix element corresponding to the right-handed neutrino exchange. We have also considered the momentum dependent contributions to NDBD  i.e., the $\lambda$ and $\eta$ contributions to NDBD. The particle physics parameter that measures the lepton number violation in case of $\lambda$ and $\eta$ contribution, are given by equations \ref{eq46} and \ref{eq47}. The effective Majorana neutrino mass due to $\lambda$  and $\eta$ contribution is thus given by,
	\begin{equation}\label{eq61}
	M^\lambda_{eff}=\frac{\eta_\lambda}{m_e} ,   \\ M^\eta_{eff}=\frac{\eta_\eta}{m_e}.
	\end{equation}

	We evaluated the half-lives for the different contributions to NDBD with respect to the elements in $M_D$ and $M_{RR}$ as well as for the total contribution using equation \ref{eq48}, for the classes A1 (NO) and B1, B2, B3, B4 (NO and IO). The half-lives corresponding to the individual contributions in the LRSM framework are shown in figures  \ref{fig13} to  \ref{fig22} and the half-life from the total contribution is shown in figure \ref{fig23} to \ref{fig27}. Apart from the light neutrino contribution, we show the individual as well as a total contribution to half-life in terms of the parameters in $M_D, M_{RR}$ for the chosen classes discussed in section \ref{sec2a}. The parameters $a1, a2, a3, b1, b2, b3, c1, c2, c3$ correspond to different entries in different chosen textures of $M_D$ while $w, x$ correspond to non-zero entries in $M_{RR}$.
	\begin{table}[h!]
		\centering
		\begin{tabular}{||c|| c|| }
			\hline
			Parameter	&Value  \\ \hline \hline
			$\gamma$	&	$10^{-9}$\\ \hline
			$\xi$	&	$10^{-6}$\\ \hline
			$M_{{\Delta_R}^{++}}\approx M_{{\Delta_L}^{++}}\approx M_{{\Delta_L}^{+}}$	&	1 TeV\\ \hline
			$M_{W_L}$	&	80 GeV\\ \hline
			$M_{W_R}$	&	4.5 TeV \\ \hline\hline		
		\end{tabular}
		\caption{The numerical values of different parameters in minimal LRSM adopted in our numerical analysis} \label{4}
	\end{table}
	
	\par
\begin{figure}[h!]
		\centering
		\includegraphics[width=0.4\textwidth]{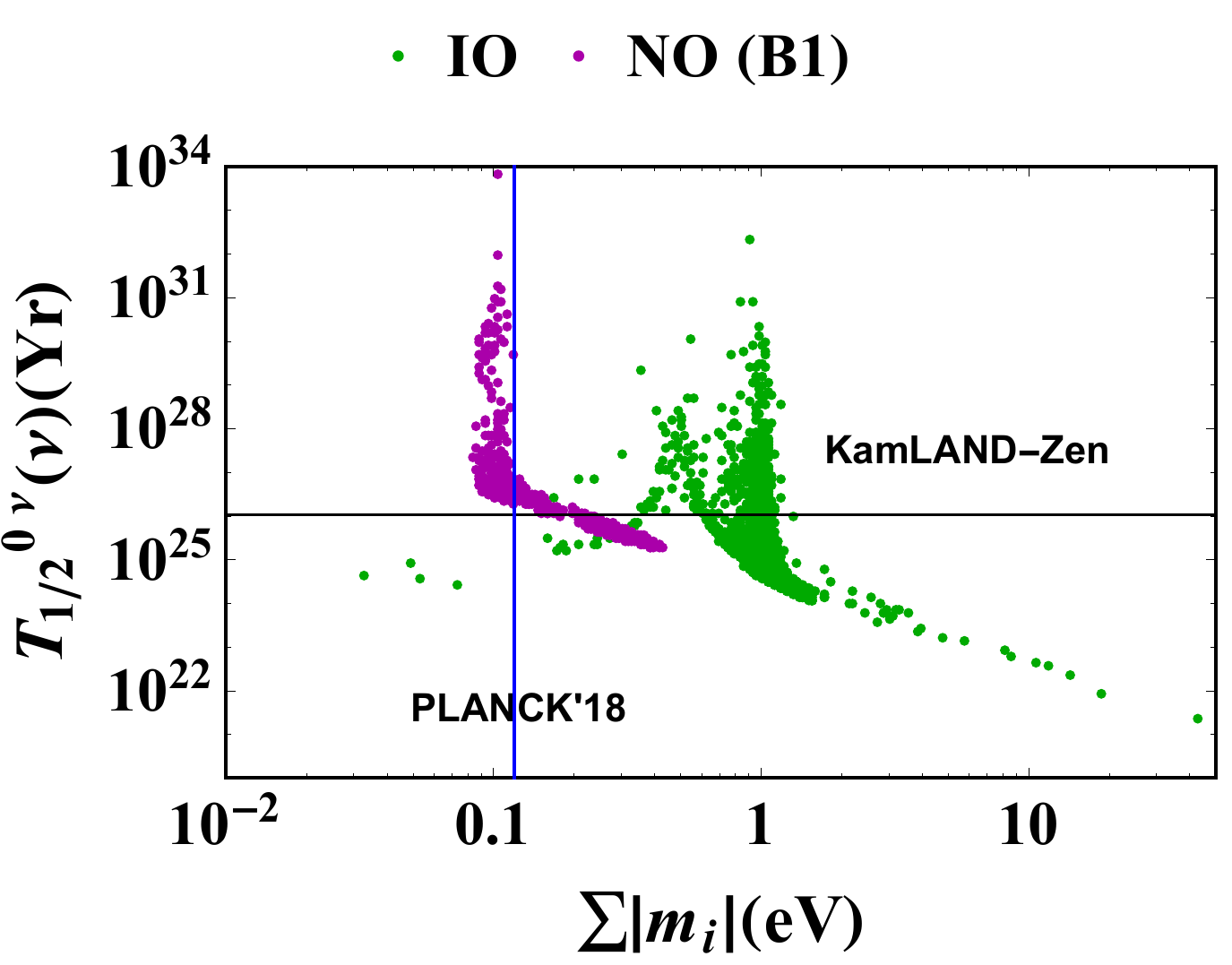}
		\includegraphics[width=0.4\textwidth]{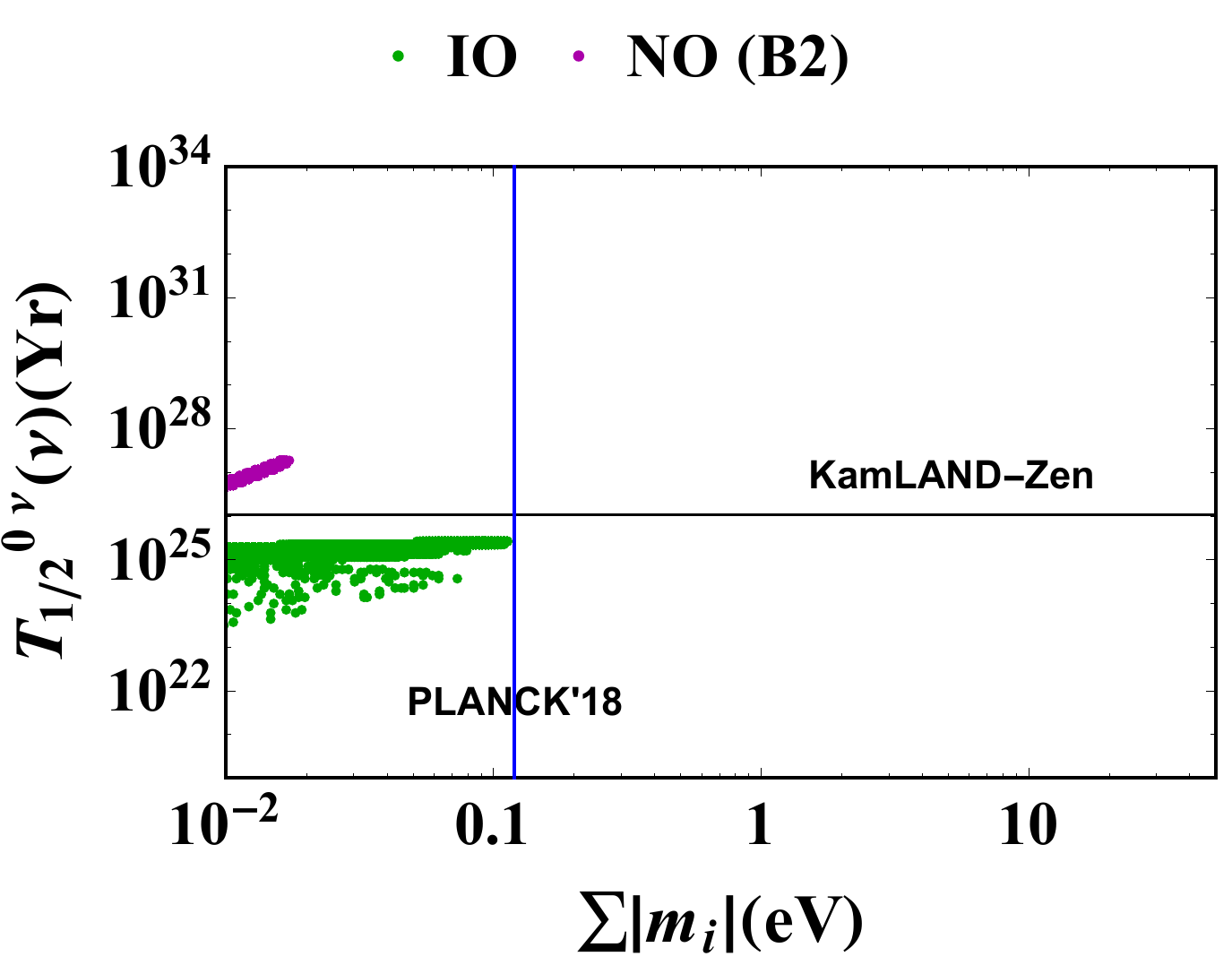} \\
		\includegraphics[width=0.4\textwidth]{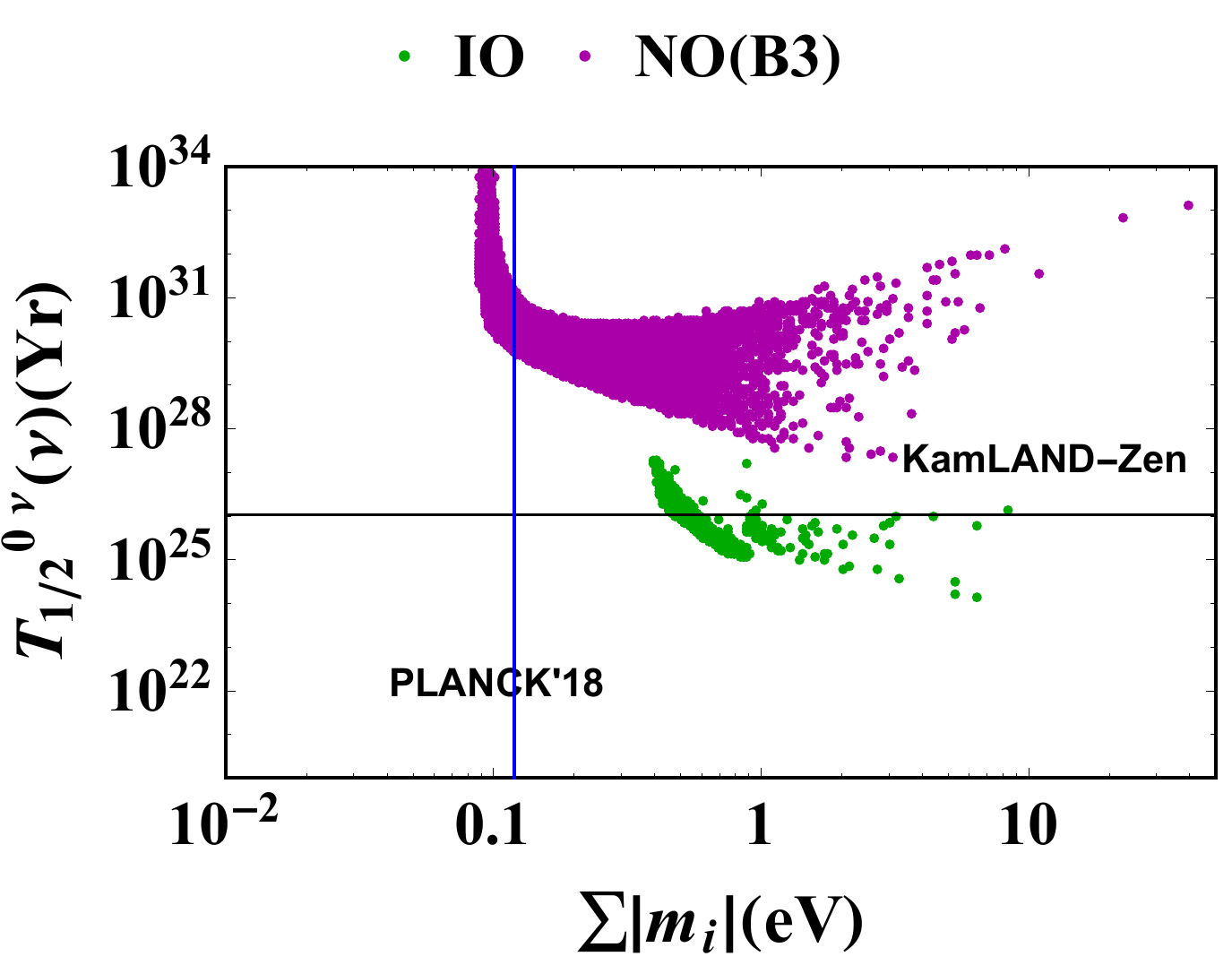}
		\includegraphics[width=0.4\textwidth]{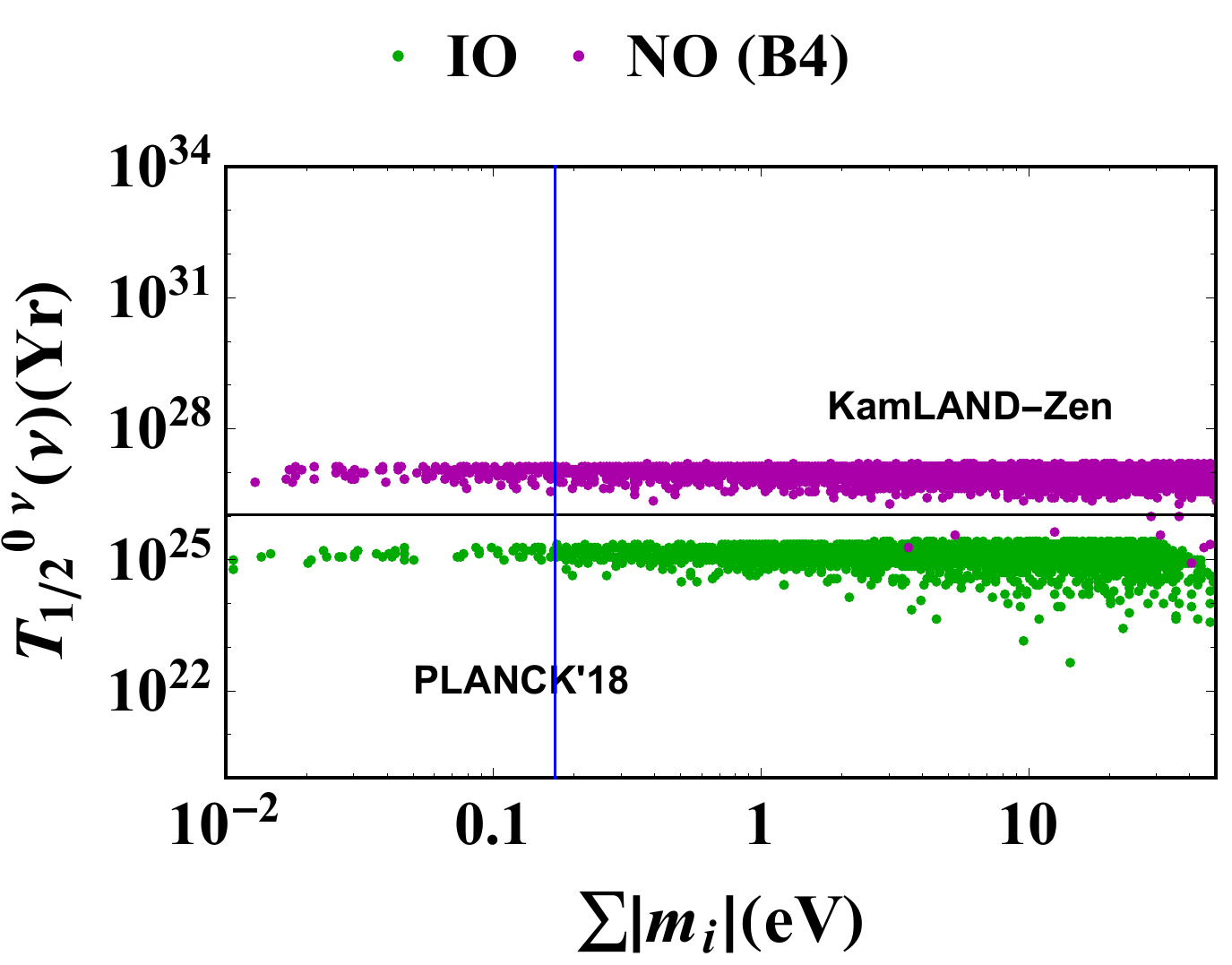}	
		\caption{ Light neutrino contribution to half-life governing NDBD as a function of the sum of light neutrino mass. The solid blue (vertical) and black (horizontal) line represents the Planck upper bound of sum of absolute neutrino mass and the KamLAND-Zen lower limit on half-life respectively.} \label{fig28}
	\end{figure}

\begin{figure}[h!]
		\includegraphics[width=0.32\textwidth,height=4cm]{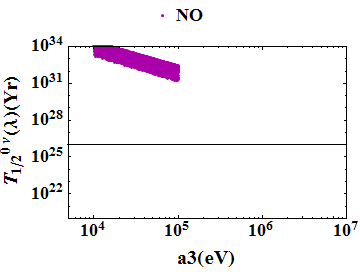}
		\includegraphics[width=0.32\textwidth,height=4cm]{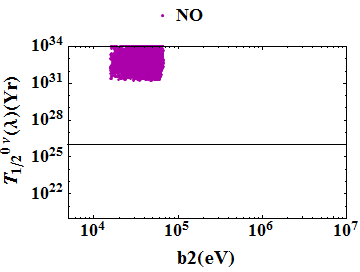}
		\includegraphics[width=0.32\textwidth,height=4cm]{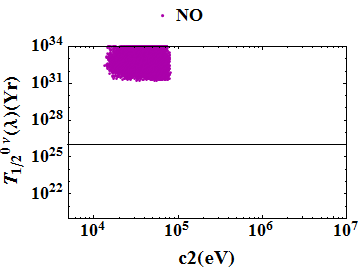}\\
		\includegraphics[width=0.32\textwidth,height=4cm]{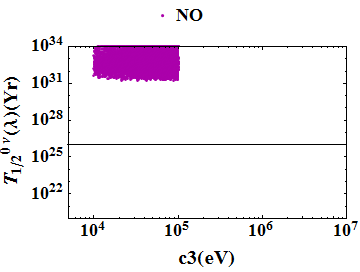}
		\includegraphics[width=0.32\textwidth,height=4cm]{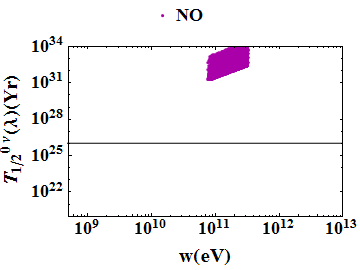}
		\includegraphics[width=0.32\textwidth,height=4cm]{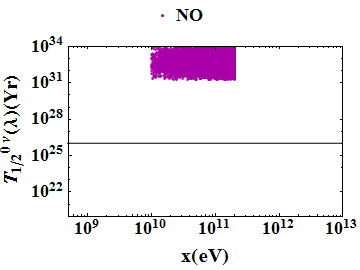}
		\caption{$\lambda$  contribution to half-life governing NDBD as a function of model  parameters for the class A1. The horizontal line represents the KamLAND-Zen lower limit.} \label{fig13}
	\end{figure}

\begin{figure}[h!]
		\includegraphics[width=0.32\textwidth,height=4cm]{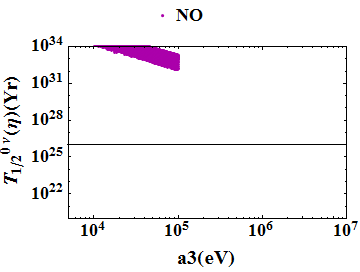}
		\includegraphics[width=0.32\textwidth,height=4cm]{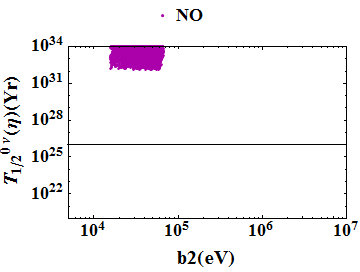}
		\includegraphics[width=0.32\textwidth,height=4cm]{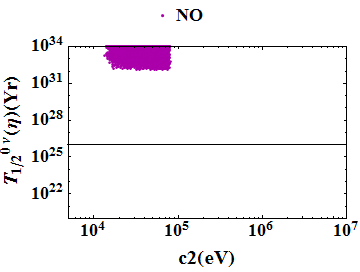}\\
		\includegraphics[width=0.32\textwidth,height=4cm]{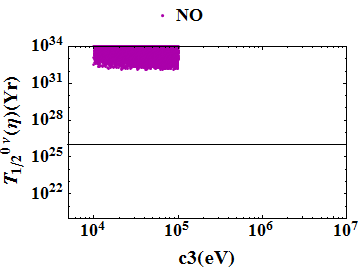}
		\includegraphics[width=0.32\textwidth,height=4cm]{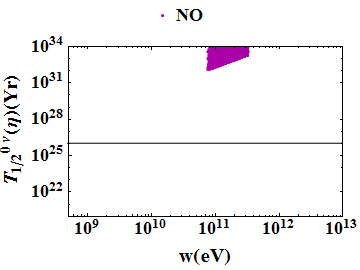}
		\includegraphics[width=0.32\textwidth,height=4cm]{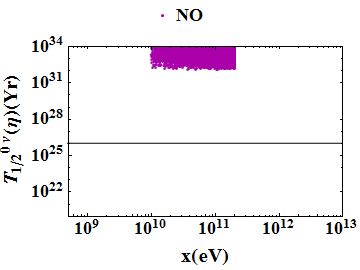}
		\caption{$\eta$  contribution to half-life governing NDBD as a function of model  parameters for the class A1. The horizontal line represents the KamLAND-Zen lower limit.} \label{fig14}
	\end{figure}

\begin{figure}[h!]
		\includegraphics[width=0.32\textwidth,height=4cm]{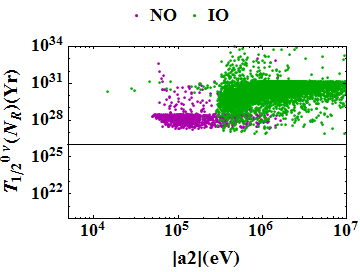}
		\includegraphics[width=0.32\textwidth,height=4cm]{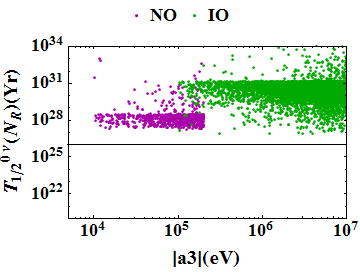}
		\includegraphics[width=0.32\textwidth,height=4cm]{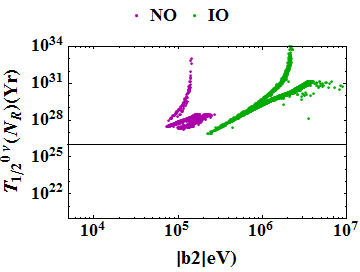}\\
		\includegraphics[width=0.32\textwidth,height=4cm]{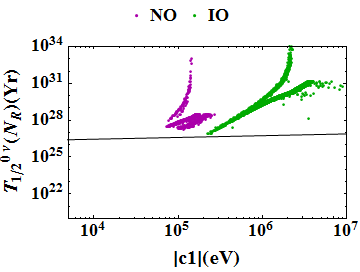}
		\includegraphics[width=0.32\textwidth,height=4cm]{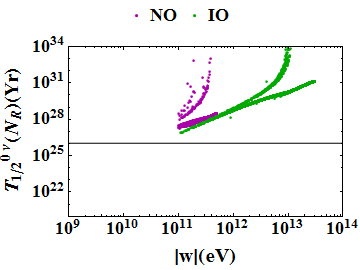}
		\includegraphics[width=0.32\textwidth,height=4cm]{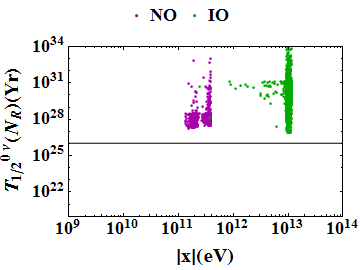}
		\caption{Heavy $\nu$ (N) contribution to half-life governing NDBD as a function of model  parameters for the class B1. The horizontal line represents the KamLAND-Zen lower limit.} \label{fig15}
	\end{figure}

\begin{figure}[h!]
		\includegraphics[width=0.32\textwidth,height=4cm]{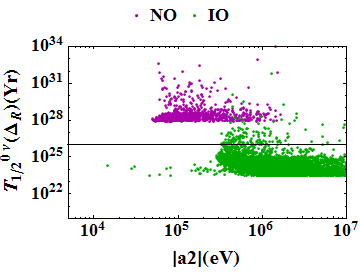}
		\includegraphics[width=0.32\textwidth,height=4cm]{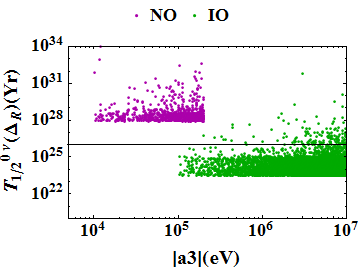}
		\includegraphics[width=0.32\textwidth,height=4cm]{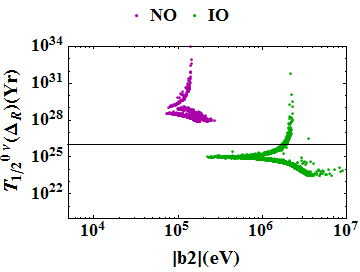}\\
		\includegraphics[width=0.32\textwidth,height=4cm]{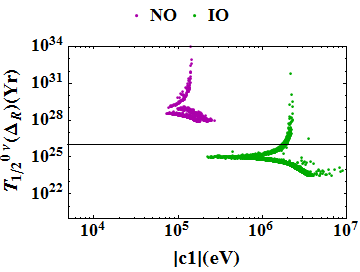}
		\includegraphics[width=0.32\textwidth,height=4cm]{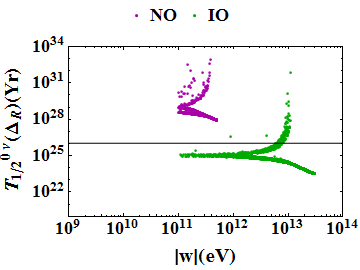}
		\includegraphics[width=0.32\textwidth,height=4cm]{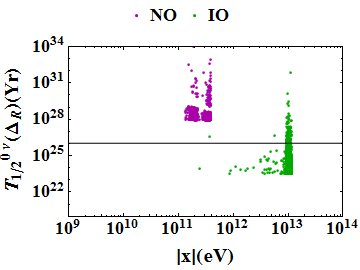}
		\caption{Heavy $\Delta_R$  contribution to half-life governing NDBD as a function of model  parameters for the class B1. The horizontal line represents the KamLAND-Zen lower limit.} \label{fig16}
	\end{figure}

\begin{figure}[h!]
		\includegraphics[width=0.32\textwidth,height=4cm]{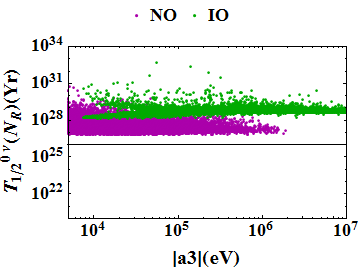}
		\includegraphics[width=0.32\textwidth,height=4cm]{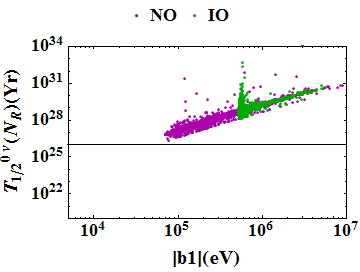}
		\includegraphics[width=0.32\textwidth,height=4cm]{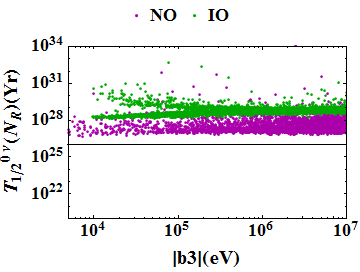}\\
		\includegraphics[width=0.32\textwidth,height=4cm]{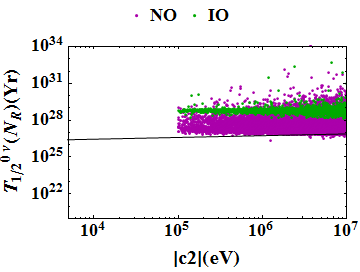}
		\includegraphics[width=0.32\textwidth,height=4cm]{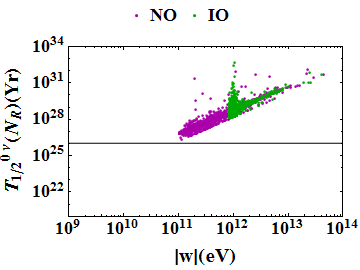}
		\includegraphics[width=0.32\textwidth,height=4cm]{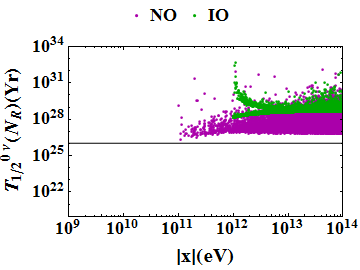}
		\caption{Heavy $\nu$ (N) contribution to half-life governing NDBD as a function of model  parameters for the class B2. The horizontal line represents the KamLAND-Zen lower limit.} \label{fig17}
	\end{figure}

\begin{figure}[h!]
		\includegraphics[width=0.32\textwidth,height=4cm]{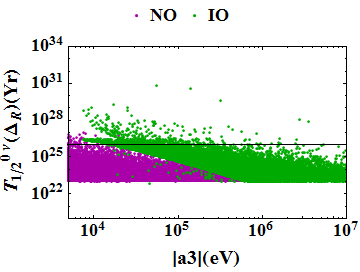}
		\includegraphics[width=0.32\textwidth,height=4cm]{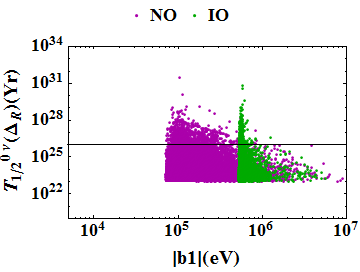}
		\includegraphics[width=0.32\textwidth,height=4cm]{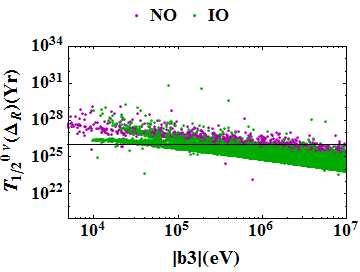}\\
		\includegraphics[width=0.32\textwidth,height=4cm]{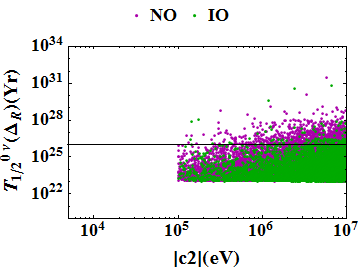}
		\includegraphics[width=0.32\textwidth,height=4cm]{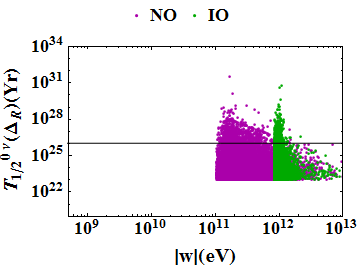}
		\includegraphics[width=0.32\textwidth,height=4cm]{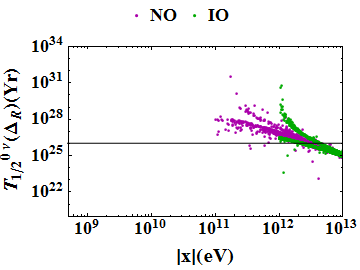}
		\caption{Heavy $\Delta_R$  contribution to half-life governing NDBD as a function of model  parameters for the class B2. The horizontal line represents the KamLAND-Zen lower limit.} \label{fig18}
	\end{figure}

\begin{figure}[h!]
		\includegraphics[width=0.32\textwidth,height=4cm]{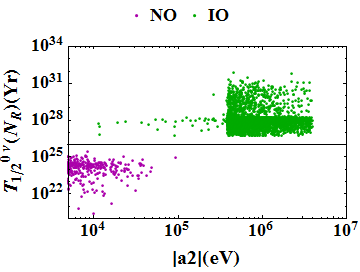}
		\includegraphics[width=0.32\textwidth,height=4cm]{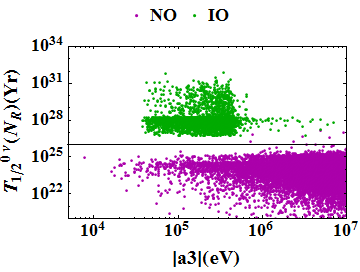}
		\includegraphics[width=0.32\textwidth,height=4cm]{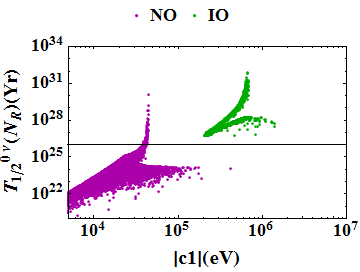}\\
		\includegraphics[width=0.32\textwidth,height=4cm]{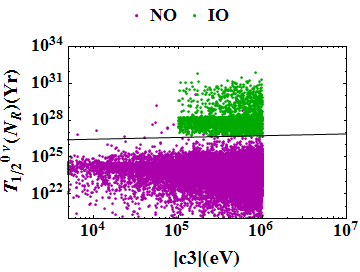}
		\includegraphics[width=0.32\textwidth,height=4cm]{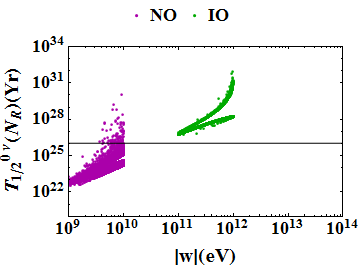}
		\includegraphics[width=0.32\textwidth,height=4cm]{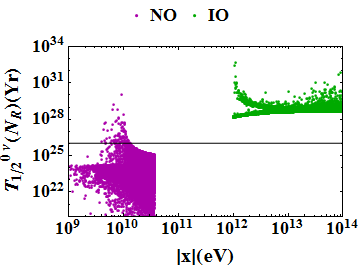}
		\caption{Heavy $\nu$ (N) contribution to half-life governing NDBD as a function of model  parameters for the class B3. The horizontal line represents the KamLAND-Zen lower limit.} \label{fig19}
	\end{figure}

\begin{figure}[h!]
		\includegraphics[width=0.32\textwidth,height=4cm]{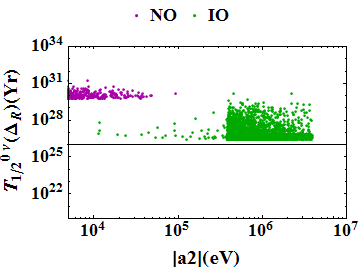}
		\includegraphics[width=0.32\textwidth,height=4cm]{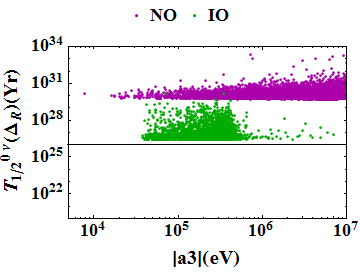}
		\includegraphics[width=0.32\textwidth,height=4cm]{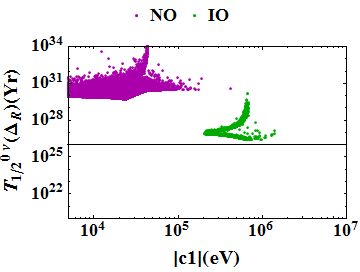}\\
		\includegraphics[width=0.32\textwidth,height=4cm]{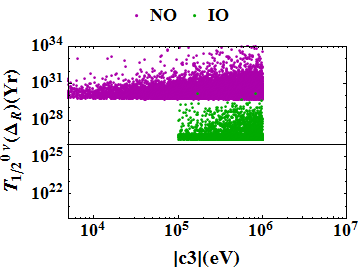}
		\includegraphics[width=0.32\textwidth,height=4cm]{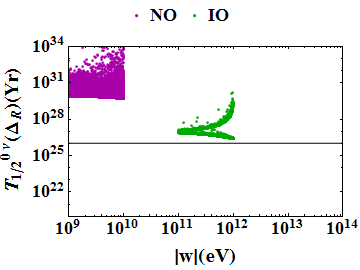}
		\includegraphics[width=0.32\textwidth,height=4cm]{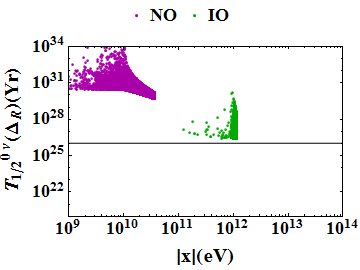}
		\caption{Heavy $\Delta_R$  contribution to half-life governing NDBD as a function of model  parameters for the class B3. The horizontal line represents the KamLAND-Zen lower limit.} \label{fig20}
	\end{figure}

\begin{figure}[h!]
		\includegraphics[width=0.32\textwidth,height=4cm]{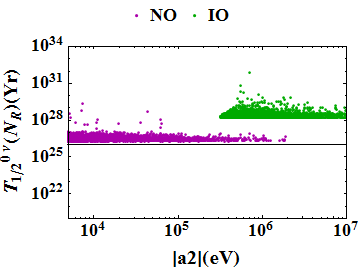}
		\includegraphics[width=0.32\textwidth,height=4cm]{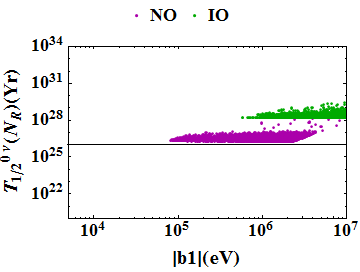}
		\includegraphics[width=0.32\textwidth,height=4cm]{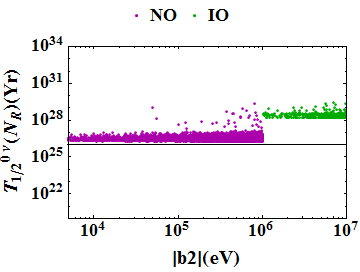}\\
		\includegraphics[width=0.32\textwidth,height=4cm]{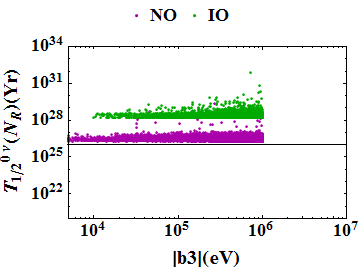}
		\includegraphics[width=0.32\textwidth,height=4cm]{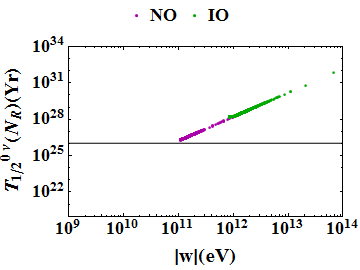}
		\includegraphics[width=0.32\textwidth,height=4cm]{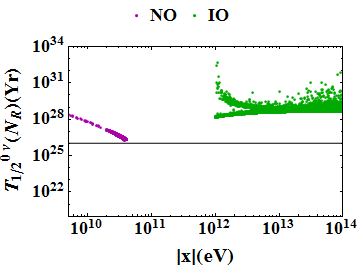}
		\caption{Heavy $\nu$ (N) contribution to half-life governing NDBD as a function of model  parameters for the class B4. The horizontal line represents the KamLAND-Zen lower limit.} \label{fig21}
	\end{figure}

\begin{figure}[h!]
		\includegraphics[width=0.32\textwidth,height=4cm]{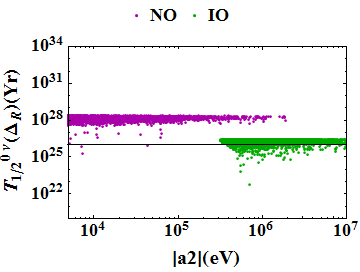}
		\includegraphics[width=0.32\textwidth,height=4cm]{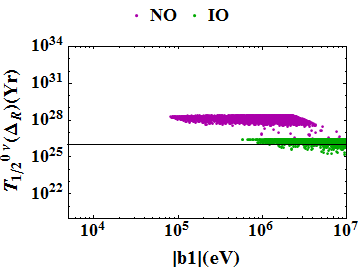}
		\includegraphics[width=0.32\textwidth,height=4cm]{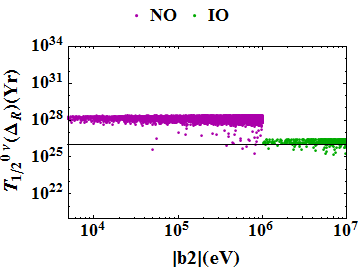}\\
		\includegraphics[width=0.32\textwidth,height=4cm]{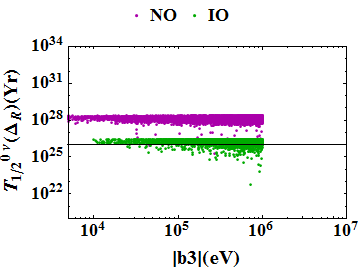}
		\includegraphics[width=0.32\textwidth,height=4cm]{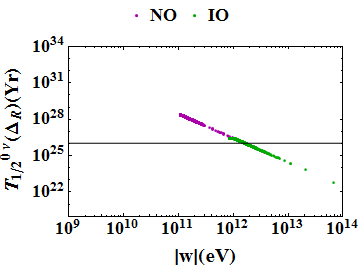}
		\includegraphics[width=0.32\textwidth,height=4cm]{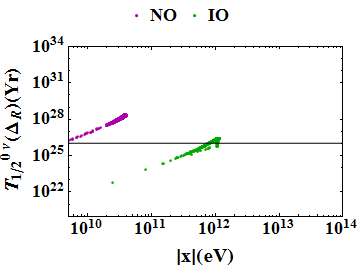}
		\caption{$\Delta_R$  contribution to half-life governing NDBD as a function of model  parameters for the class B4. The horizontal line represents the KamLAND-Zen lower limit.} \label{fig22}
	\end{figure}

\begin{figure}[h!]
		\includegraphics[width=0.32\textwidth,height=4cm]{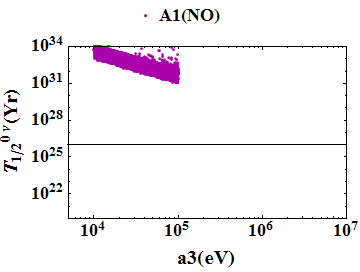}
		\includegraphics[width=0.32\textwidth,height=4cm]{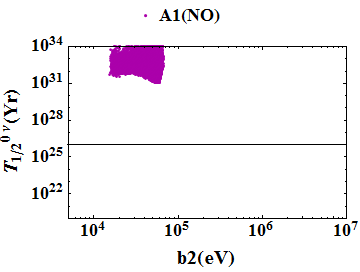}
		\includegraphics[width=0.32\textwidth,height=4cm]{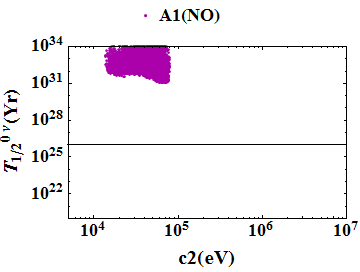}\\
		\includegraphics[width=0.32\textwidth,height=4cm]{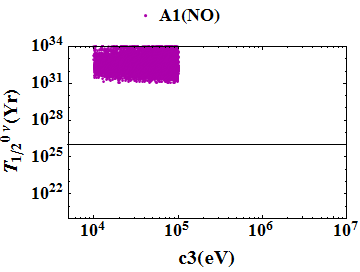}
		\includegraphics[width=0.32\textwidth,height=4cm]{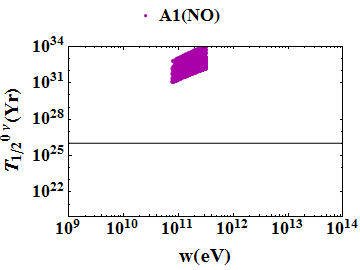}
		\includegraphics[width=0.32\textwidth,height=4cm]{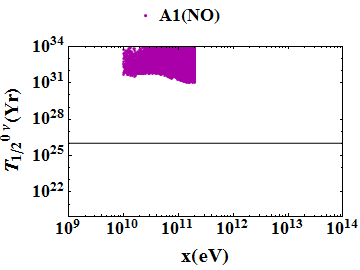}
		
		\caption{ Total  contribution to half-life governing NDBD as a function of the model parameters for the class A1. The horizontal line represents the KamLAND-Zen lower limit } \label{fig23}
	\end{figure}

\begin{figure}[h!]
		\includegraphics[width=0.32\textwidth,height=4cm]{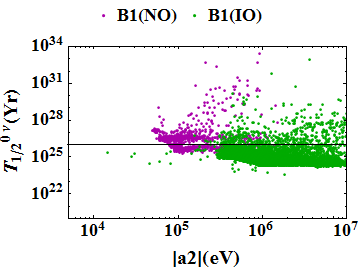}
		\includegraphics[width=0.32\textwidth,height=4cm]{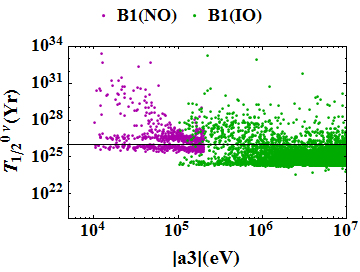}
		\includegraphics[width=0.32\textwidth,height=4cm]{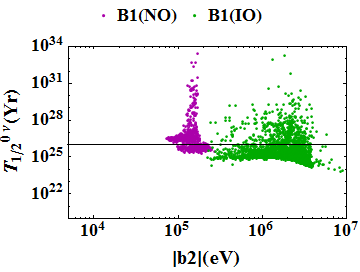}\\
		\includegraphics[width=0.32\textwidth,height=4cm]{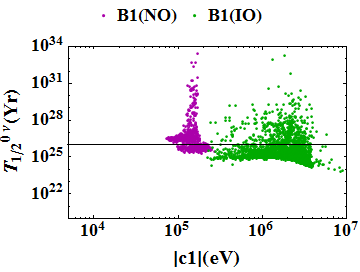}
		\includegraphics[width=0.32\textwidth,height=4cm]{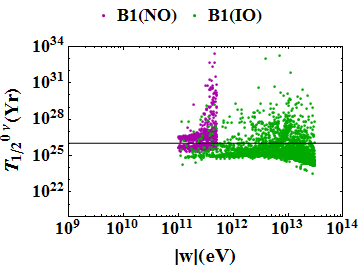}
		\includegraphics[width=0.32\textwidth,height=4cm]{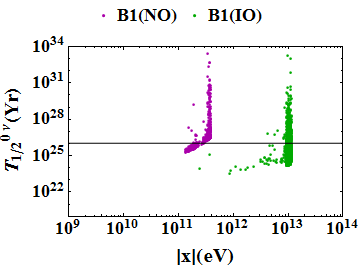}
		
		\caption{ Total  contribution to half-life governing NDBD as a function of the model parameters for the class B1. The horizontal line represents the KamLAND-Zen lower limit.} \label{fig24}
	\end{figure}

\begin{figure}[h!]
		\includegraphics[width=0.32\textwidth,height=4cm]{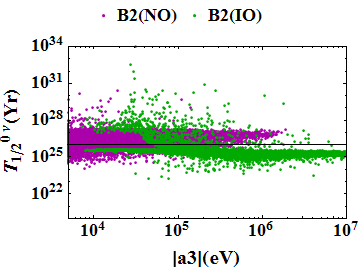}
		\includegraphics[width=0.32\textwidth,height=4cm]{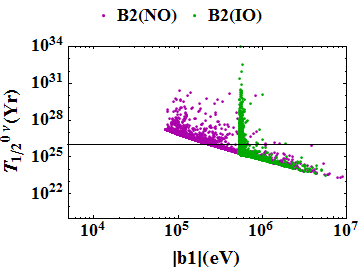}
		\includegraphics[width=0.32\textwidth,height=4cm]{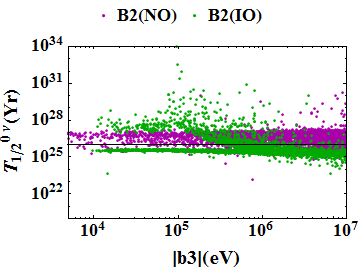}\\
		\includegraphics[width=0.32\textwidth,height=4cm]{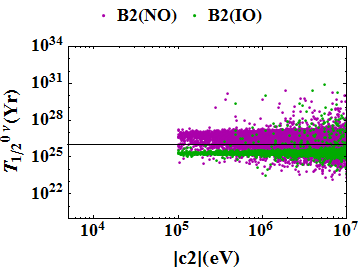}
		\includegraphics[width=0.32\textwidth,height=4cm]{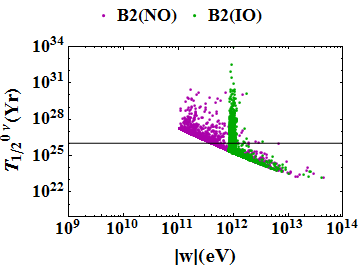}
		\includegraphics[width=0.32\textwidth,height=4cm]{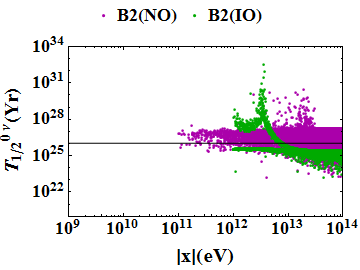}
		
		\caption{ Total  contribution to half-life governing NDBD as a function of the model parameters for the class B2. The horizontal line represents the KamLAND-Zen lower limit.} \label{fig25}
	\end{figure}

\begin{figure}[h!]
		\includegraphics[width=0.32\textwidth,height=4cm]{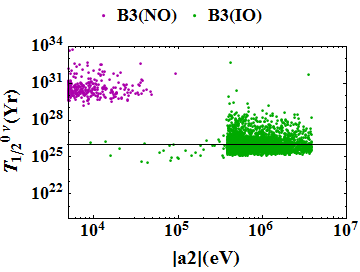}
		\includegraphics[width=0.32\textwidth,height=4cm]{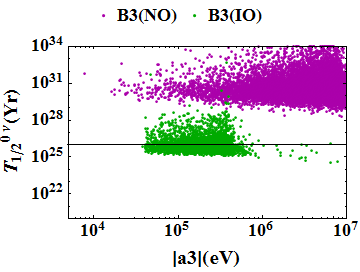}
		\includegraphics[width=0.32\textwidth,height=4cm]{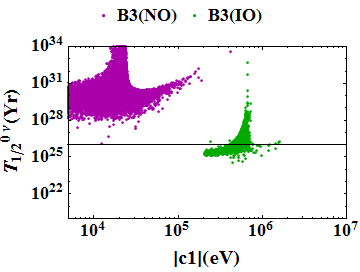}\\
		\includegraphics[width=0.32\textwidth,height=4cm]{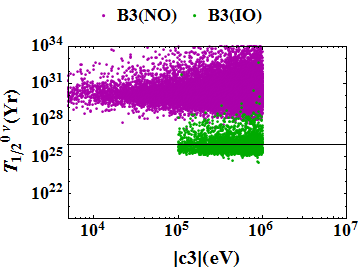}
		\includegraphics[width=0.32\textwidth,height=4cm]{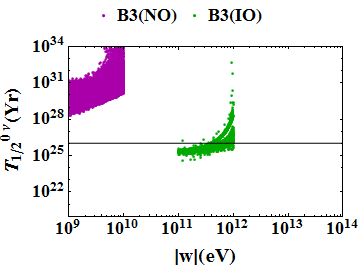}
		\includegraphics[width=0.32\textwidth,height=4cm]{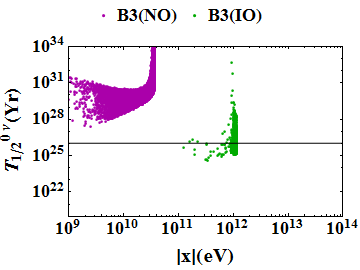}
		
		\caption{ Total  contribution to half-life governing NDBD as a function of the model parameters for the class B3. The horizontal line represents the KamLAND-Zen lower limit.} \label{fig26}
	\end{figure}

\begin{figure}[h!]
		\includegraphics[width=0.32\textwidth,height=4cm]{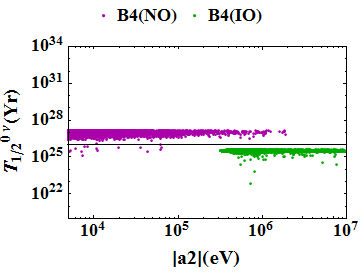}
		\includegraphics[width=0.32\textwidth,height=4cm]{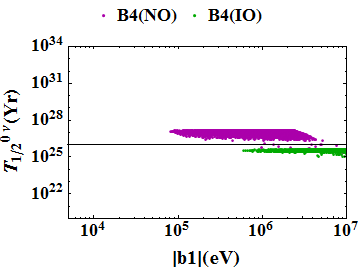}
		\includegraphics[width=0.32\textwidth,height=4cm]{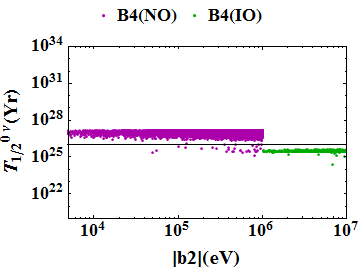}\\
		\includegraphics[width=0.32\textwidth,height=4cm]{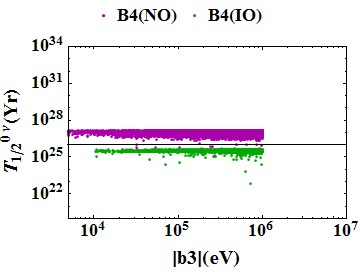}
		\includegraphics[width=0.32\textwidth,height=4cm]{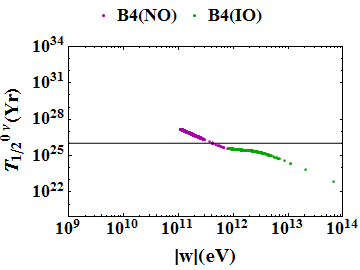}
		\includegraphics[width=0.32\textwidth,height=4cm]{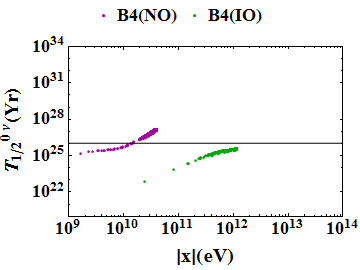}
		
		\caption{ Total  contribution to half-life governing NDBD as a function of the model parameters for the class B4. The horizontal line represents the KamLAND-Zen lower limit.} \label{fig27}
	\end{figure}

\begin{figure}[h!]
		\centering
		\includegraphics[width=0.32\textwidth,height=4cm]{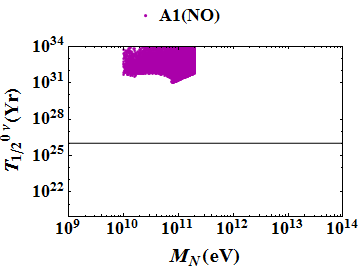}
		\includegraphics[width=0.32\textwidth,height=4cm]{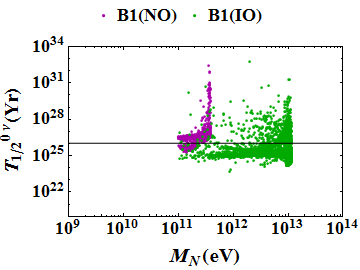}
		\includegraphics[width=0.32\textwidth,height=4cm]{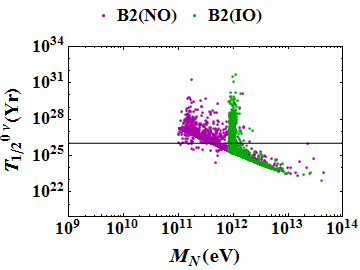}\\
		\includegraphics[width=0.32\textwidth,height=4cm]{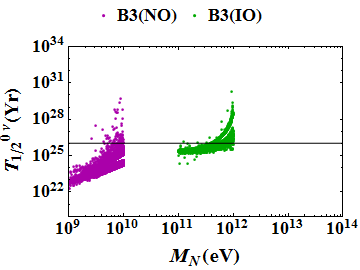}
		\includegraphics[width=0.32\textwidth,height=4cm]{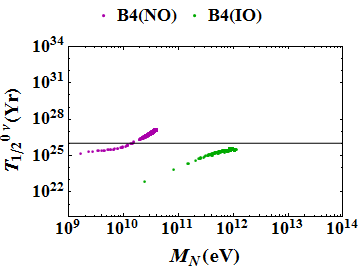}	
		\caption{Total  contribution to half-life governing NDBD as a function of the lightest right handed neutrino mass.} \label{fig29}
	\end{figure}

 From figure \ref{fig1} to \ref{fig3}, we have shown the correlation between different neutrino parameters in the framework of LRSM for both normal and inverted hierarchies. Figures \ref{fig13} and \ref{fig14} represent the half-life governing NDBD for different individual contributions in LRSM for the class A1. Furthermore, due to $\left[{M_\nu}\right]_{ee}=0$, the standard light
	neutrino contribution does not arise in this case. Again, as seen in equation \ref{eq37}, $\left[{M_{RR}}\right]_{ee}=0$ for the class A1, so the heavy neutrino contributions mediated by right-handed currents also cease to exist in this case. Due to the inconsistency of IO with experimental data, we have analysed only for the normal case. In figures \ref{fig15} and \ref{fig16}, we have shown the half-life for the class B1 for both the mass hierarchies. However, it is seen that the mixed contributions do not arise in this case as the factor governing NDBD for the left-right mixing, $\left[M_D M_{RR}^{-1}\right]_{ee}$ is almost negligible in this case. Similar results hold for the classes B2, B3 and B4. For the classes, B2, B3 and B4 we have shown the individual contributions in figures \ref{fig17} to \ref{fig18}, \ref{fig19} to \ref{fig20}, \ref{fig21} to \ref{fig22} respectively. We have also shown the total contributions to NDBD in LRSM scenario in the figures \ref{fig23} to \ref{fig27} for the different allowed classes of two zero texture neutrino mass. In all the classes, we have varied the half-life governing NDBD with the parameters in the Dirac and Majorana mass matrix $M_D$ and $M_{RR}$ and compared with the experimental lower limit provided by the KamLAND-Zen experiment \cite{KamLAND-Zen:2016pfg}. In figure  \ref{fig28} we have shown the standard light neutrino contribution to half-life as a function of the sum of the absolute neutrino masses considering the PLANCK bound $\sum_i \lvert m_i \rvert <0.12 $ eV \cite{Aghanim:2018eyx}. From the figures, we can conclude that only NO satisfies the experimental bounds for all the classes, B1-B4. In figure \ref{fig29}, we plotted the total contribution to NDBD with the lightest right-handed neutrino mass with a view to seeing the parameter space of the heavy RH neutrino mass satisfying NDBD. 
	
\begin{figure}[h!]
		\includegraphics[width=0.32\textwidth,height=4cm]{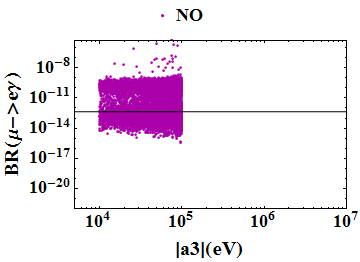}
		\includegraphics[width=0.32\textwidth,height=4cm]{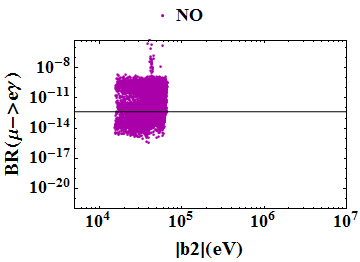}
		\includegraphics[width=0.32\textwidth,height=4cm]{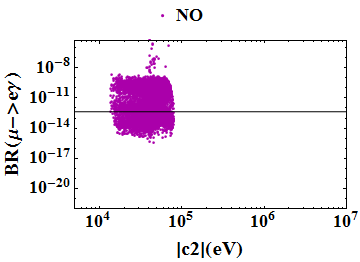}\\
		\includegraphics[width=0.32\textwidth,height=4cm]{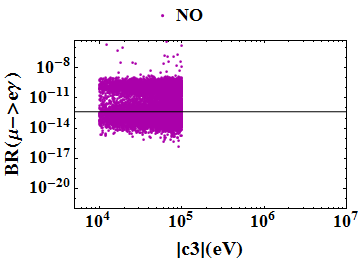}
		\includegraphics[width=0.32\textwidth,height=4cm]{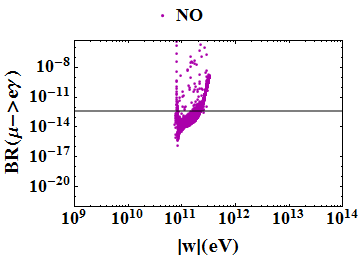}
		\includegraphics[width=0.32\textwidth,height=4cm]{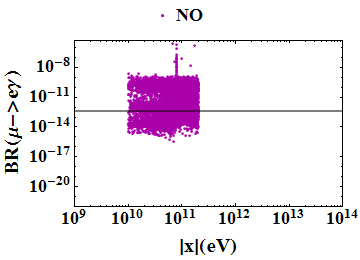}
		
		\caption{  BR for $\rm \mu\rightarrow e\gamma $ as a function of model  parameters for the class A1. The horizontal line represents the upper limit for BR given by MEG experiment.}\label{fig30}
	\end{figure}

\begin{figure}[h!]
		\includegraphics[width=0.32\textwidth,height=4cm]{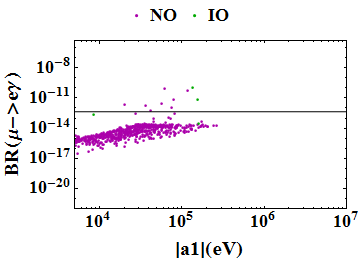}
		\includegraphics[width=0.32\textwidth,height=4cm]{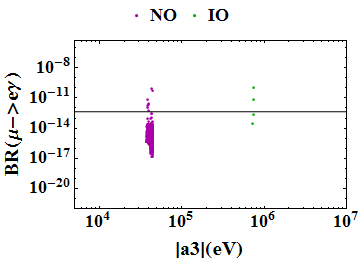}
		\includegraphics[width=0.32\textwidth,height=4cm]{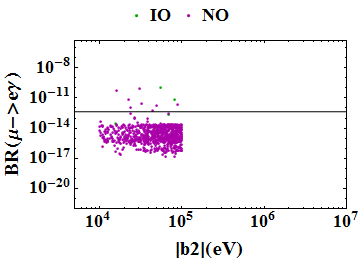}\\
		\includegraphics[width=0.32\textwidth,height=4cm]{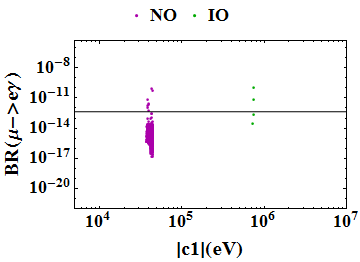}
		\includegraphics[width=0.32\textwidth,height=4cm]{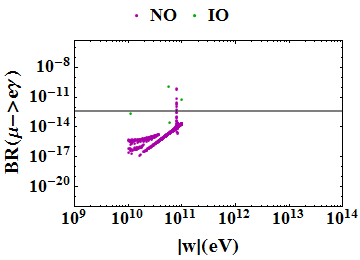}
		\includegraphics[width=0.32\textwidth,height=4cm]{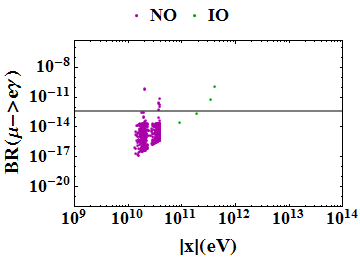}
		
		\caption{  BR for $\rm \mu\rightarrow e\gamma $ as a function of model  parameters for the class B1. The horizontal line represents the upper limit for BR given by MEG experiment.}\label{fig31}
	\end{figure}

\begin{figure}[h!]
		\includegraphics[width=0.32\textwidth,height=4cm]{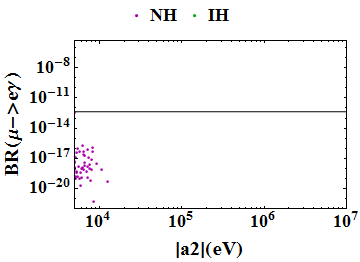}
		\includegraphics[width=0.32\textwidth,height=4cm]{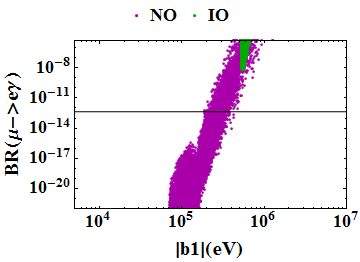}
		\includegraphics[width=0.32\textwidth,height=4cm]{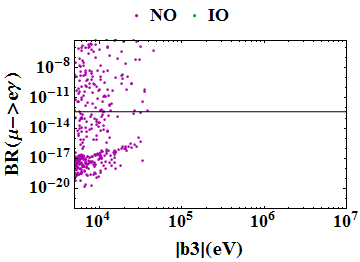}\\
		\includegraphics[width=0.32\textwidth,height=4cm]{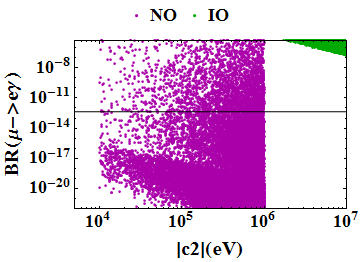}
		\includegraphics[width=0.32\textwidth,height=4cm]{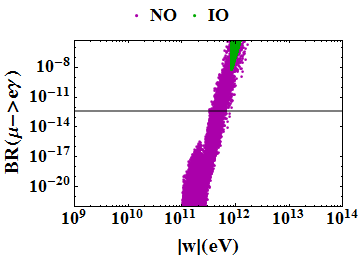}
		\includegraphics[width=0.32\textwidth,height=4cm]{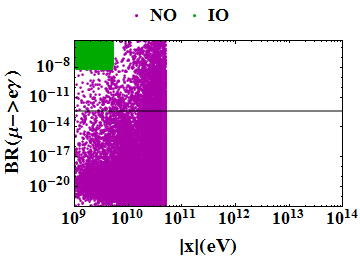}
		
		\caption{  BR for $\rm \mu\rightarrow e\gamma $ as a function of model  parameters for the class B2. The horizontal line represents the upper limit for BR given by MEG experiment.}\label{fig32}
	\end{figure}

\begin{figure}[h!]
		\includegraphics[width=0.32\textwidth,height=4cm]{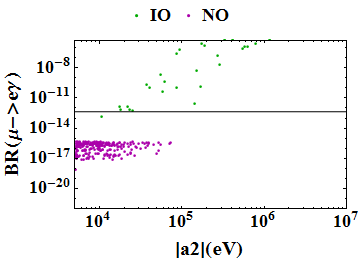}
		\includegraphics[width=0.32\textwidth,height=4cm]{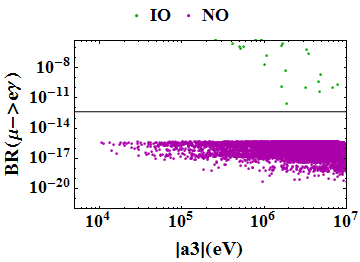}
		\includegraphics[width=0.32\textwidth,height=4cm]{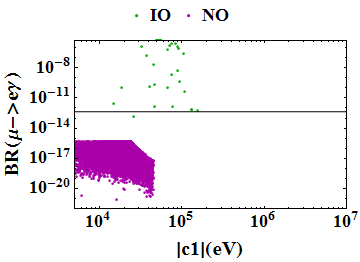}\\
		\includegraphics[width=0.32\textwidth,height=4cm]{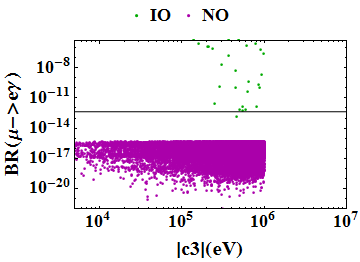}
		\includegraphics[width=0.32\textwidth,height=4cm]{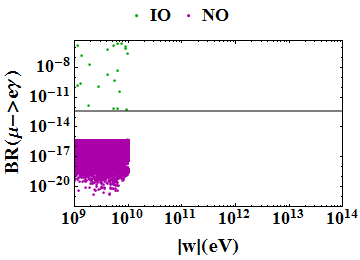}
		\includegraphics[width=0.32\textwidth,height=4cm]{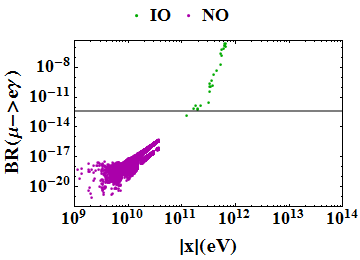}
		
		\caption{  BR for $\rm \mu\rightarrow e\gamma $ as a function of model  parameters for the class B3. The horizontal line represents the upper limit for BR given by MEG experiment.}\label{fig33}
		
	\end{figure}

\begin{figure}[h!]
		\includegraphics[width=0.32\textwidth,height=4cm]{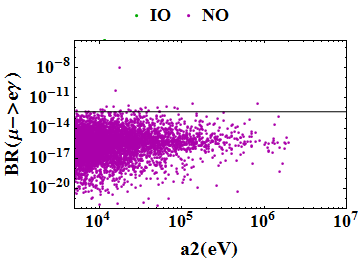}
		\includegraphics[width=0.32\textwidth,height=4cm]{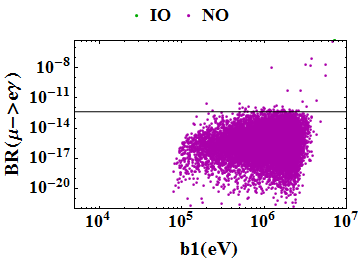}
		\includegraphics[width=0.32\textwidth,height=4cm]{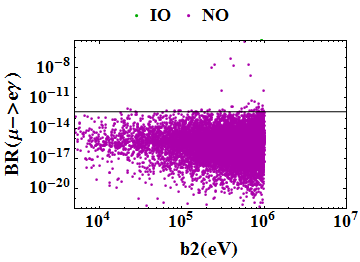}\\
		\includegraphics[width=0.32\textwidth,height=4cm]{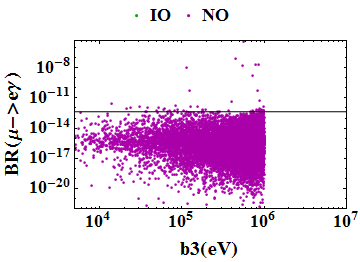}
		\includegraphics[width=0.32\textwidth,height=4cm]{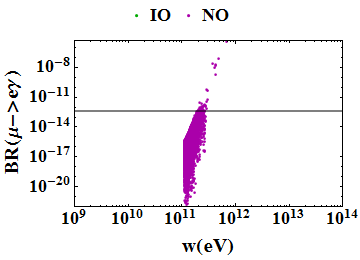}
		\includegraphics[width=0.32\textwidth,height=4cm]{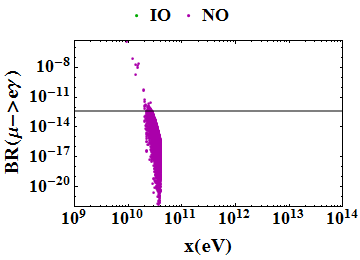}
		
		\caption{  BR for $\rm \mu\rightarrow e\gamma $ as a function of model  parameters for the class B4. The horizontal line represents the upper limit for BR given by MEG experiment.}\label{fig34}
	\end{figure}

\begin{figure}[h!]
		\includegraphics[width=0.32\textwidth,height=4cm]{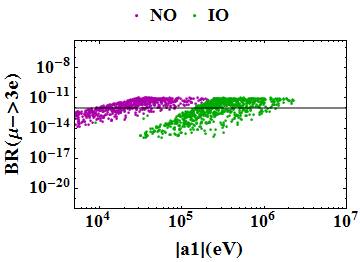}
		\includegraphics[width=0.32\textwidth,height=4cm]{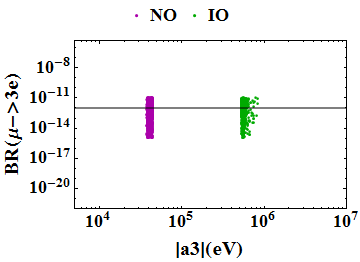}
		\includegraphics[width=0.32\textwidth,height=4cm]{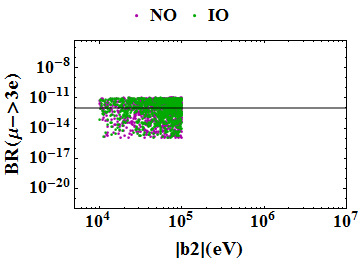}\\
		\includegraphics[width=0.32\textwidth,height=4cm]{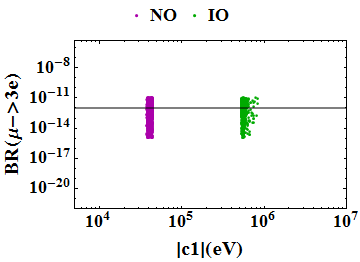}
		\includegraphics[width=0.32\textwidth,height=4cm]{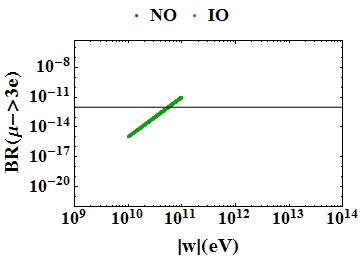}
		\includegraphics[width=0.32\textwidth,height=4cm]{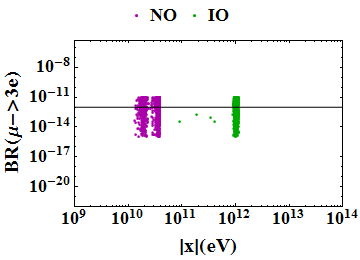}
		
		\caption{  BR for $\rm \mu\rightarrow 3e $ as a function of model  parameters for the class B1. The horizontal line represents the upper limit for BR given by SINDRUM experiment.}\label{fig36}
	\end{figure}

	Furthermore, we have also evaluated the branching ratio of the CLFV process $\mu\rightarrow e\gamma $ with respect to the elements of $M_D$ and $M_{RR}$ for the different classes of two zero texture neutrino mass for both normal and inverted hierarchies.
	For calculating the BR, we used the expression given in equation (\ref{eq49}). The relevant calculations were done by diagonalizing the right- handed neutrino mass matrix and obtaining the mixing matrix element, $V_{ij}$ and the eigenvalues $M_i$. The results obtained have been summarized in the figures \ref{fig30} to \ref{fig34} where the BR is plotted as a function of parameters in $M_D, M_{RR}$, along with the comparison with MEG upper bound. Furthermore, we have also studied the BR for the LFV process $\left(\mu\rightarrow 3e\right)$ and show the results in figure \ref{fig36} with the parameters in $M_D$ and $M_{RR}$ for the different classes and compared with the experimental upper bound provided by the SINDRUM experiment. The BR for both the CLFV processes have strong dependence on the right-handed neutrino mixing matrix structure. Interestingly, we see that IO occupies very less parameter space within experimental bound in comparison to NO. For the class B4, all the parameter space is ruled out by MEG upper limit. For the process $\left(\mu\rightarrow 3e\right)$, the BR is controlled by $\rm h'_{\mu e}{h'_{ee}}^{*}$ which vanishes for the classes A1, B2, B3 and B4 due to vanishing $\rm h'_{\mu e}$ because of the structure of $M_{RR}$. Whereas for the classes B1, using the structures of $M_D$ and $M_{RR}$ as shown in equation \ref{eq39b} we arrive at the BR as shown in figure  \ref{fig36}. Again, we can see from our analysis that the observables for NDBD and CLFV are highly dependent on the Dirac and Majorana mass matrices and their structures which are again different for the different classes of the two zero texture light neutrino mass matrix. 
	
	It is worth mentioning that several earlier works \cite{Tello:2010am, Bambhaniya:2015ipg} found that the NDBD and CLFV limits induce a hierarchy between the mass of the
	$SU(2)_R$ scalar bosons and the mass of the heaviest right-handed neutrino that
	must be 2 to 10 times smaller for
	$M_{W_R}=3.5$~TeV. These bounds are however derived under the assumption that light neutrino mass arises from either a type I or a type II seesaw mechanism. Considering a scenario with a combination of type I and type II seesaw mechanisms (as in this work) enables us to evade those bounds, as also pointed out earlier by \cite{Borah:2015ufa, Borah:2016iqd}. The $SU(2)_R$ triplet scalar masses are allowed to be even smaller than the heaviest right-handed neutrino mass. Right-handed neutrinos could nevertheless be indirectly constrained by neutrinoless double-beta decays and cosmology~\cite{Borah:2016lrl, Frank:2017tsm,
		Araz:2017qcs}.
	
The constraints from NDBD and CLFV can be complementary to the collider bounds on LRSM, as pointed out by several works including \cite{Das:2012ii, Lindner:2016lpp}. For example, NDBD constraints can rule out some part of the parameter space in the plane of the lightest right-handed neutrino mass $M_N$ and $W_R$ mass where the LHC limits \cite{Sirunyan:2018pom} are weak. As can be seen from the plots of figure \ref{fig29}, NDBD constraints can rule out lightest right-handed neutrino mass as low as 1 GeV, which remains allowed from LHC limits on same sign dilepton searches \cite{Sirunyan:2018pom}. This also agrees with the estimates derived in the earlier works mentioned above. In another recent work \cite{Lindner:2016lxq}, prospects of probing the $M_N-M_{W_R}$ plane to a much wider extent at several experiments including future colliders and future NDBD experiments were considered. Even in these studies, the collider and NDBD sensitivities were found to be complementary with NDBD experiments putting stronger limits on low $M_N \leq \mathcal{O}(10 \; \rm GeV)$ while colliders can probe high mass region $M_N \sim \mathcal{O}(\rm TeV)$.

	\clearpage
	
	\clearpage
	\clearpage
	
	\clearpage
	

	
	

	

	
	\clearpage
	
	\begin{table}[h!]
		\centering
		\begin{tabular}{|c||  c| c| c| c|| }
			\hline
			Class&	NDBD (Total half-life)  & BR($\rm \mu\rightarrow e\gamma $)  & BR($\rm \mu\rightarrow 3e$)\\ \hline
			A1(NO/IO)&$ \checkmark (\times)$	&$ \checkmark (\times)$ &$ \checkmark (\checkmark)$  \\ \hline\hline\hline
			B1(NO/IO)&$ \checkmark (\checkmark)$ 	&$ \checkmark (\checkmark)$  &$ \checkmark (\checkmark)$\\ \hline\hline\hline
			B2(NO/IO)&$ \checkmark (\checkmark)$ 	&$ \checkmark(\times)$ &$ \checkmark(\checkmark)$ \\ \hline\hline\hline
			B3(NO/IO)&$ \checkmark (\checkmark)$ 	&$ \checkmark (\checkmark)$   &$ \checkmark(\checkmark)$ \\ \hline\hline\hline
			B4(NO/IO)&$ \checkmark (\times)$   &$\checkmark (\times)$ &$ \checkmark(\checkmark)$\\ \hline		
		\end{tabular}
		\caption{ Summary of allowed and disallowed textures. The $\checkmark$ and $\times$ symbol are used to denote if the observables (NDBD/CLFV) are (not are) within the current experimental upper limit.} \label{5}
	\end{table}
	
	\begin{table}[h!]
		\centering
		\begin{tabular}{|c||c | c| c| c| c| c| c| c|| }
			\hline
			Class&$\nu_L$&  ${N_{R}}^R$ &${N_{R}}^L$&	$\Delta_R$  & $\lambda$   & $\eta$& BR($\rm \mu\rightarrow e\gamma $)&BR($\rm \mu\rightarrow 3e$)\\ \hline
			A1&& && 	& & & NO&	 \\ \hline\hline\hline
			B1&NO(IO)&NO(IO)&&NO(IO)  	&&&NO(IO)&NO(IO) \\ \hline\hline\hline
			B2&NO&NO(IO)&NO(IO)& NO(IO) 	& & &NO&\\ \hline\hline\hline
			B3&NO(IO)&IO &&NO(IO)	& & &NO(IO)&\\ \hline\hline\hline
			B4&NO&NO(IO)&& NO(IO)  &&&NO&\\ \hline
		\end{tabular}
		\caption{Summarised form of the results only for the allowed cases pointing out the individual contributions to NDBD as well as the total CLFV contributions which can saturate corresponding experimental upper limits for both NO and IO. The empty boxes correspond to the contributions which remain subdominant.} \label{6}
	\end{table}

	\section{Conclusion}{\label{sec5}}
	We have studied the possibility of texture zeros in lepton mass matrices of the minimal left-right symmetric model where light neutrino mass arises from a combination of type I and type II seesaw mechanism. Considering the allowed texture zeros in light neutrino mass matrix, we list out all possible texture zero possibilities in Dirac and heavy neutrino mass matrices which play a role in type I and type II seesaw mechanism. After making this exhaustive list in table \ref{1}, we consider, for our numerical studies, the possibility with the maximum allowed zeros in $M_{\nu}$, $M_D$ and $M_{RR}$ while keeping the rank of the latter three. After finding the allowed parameter space for two zero textures in light neutrino mass matrix $M_{\nu}$, we then evaluate the elements of $M_D, M_{RR}$ by choosing an optimistic $M_{W_R} = 4.5$ TeV while keeping the right-handed neutrino masses above 1 GeV. We then evaluate the contributions to NDBD half-life as well as CLFV decays $\mu \rightarrow e \gamma, \mu \rightarrow 3e$ and constrain the texture zero mass matrices from the relevant experimental bounds. The summary of our results is shown in table \ref{5}. It is seen that out of all the cases considered with 5-0 $M_D$ and $4-0$ $M_{RR}$, only A1 (NO), B1 (NO/IO), B2 (NO), B3 (NO/IO), B4 (NO) are allowed from both NDBD and CLFV constraints while the others are disallowed by at least one of the constraints. In table \ref{6}, we further show the allowed cases pointing out the individual contributions to NDBD and total contributions to CLFV which can saturate the current experimental upper bound, keeping them sensitive to ongoing and future experiments. It is interesting to note that even for the most conservative lower bound on left-right symmetry scale that is $M_{W_R} = 4.5$ TeV from collider experiment, the complementary bounds from rare decay experiments can rule out several texture possibilities while keeping the allowed ones sensitive to upcoming experiments. We performed our study from a phenomenological point of view keeping the framework as minimal as the minimal LRSM. We leave a more detailed study of these interesting texture zero scenarios within additional flavour symmetry for an upcoming work.
	\acknowledgments
	DB acknowledges the support from Indian Institute of Technology Guwahati start-up grant (reference number: xPHYSUGI-ITG01152xxDB001), Early Career Research Award from Science and Engineering Research Board (SERB), Department of Science and Technology (DST), Government of India (reference number: ECR/2017/001873) and Associateship Programme of IUCAA, Pune. The work of MKD is supported by the Department of Science and Technology, Government of India under the project number EMR/2017/001436.
	
\bibliographystyle{JHEP}
\bibliography{ndbdlrsm}

\end{document}